\documentclass[usenatbib]{mn2e}

\usepackage[dvips]{graphicx}
\usepackage{amssymb}
\usepackage{txfonts}
\usepackage{colordvi}

\newcommand{\fermi}{{\textit{Fermi}}}
\newcommand{\agile}{{\textit{AGILE}}}

\newcommand{\g}{$\gamma$}

\let\oldhat\hat
\renewcommand{\vec}[1]{\mathbfit{#1}}
\renewcommand{\hat}[1]{\oldhat{\mathbfit{#1}}}

\newbox\grsign \setbox\grsign=\hbox{$>$} \newdimen\grdimen \grdimen=\ht\grsign
\newbox\simlessbox \newbox\simgreatbox \newbox\simpropbox
\setbox\simgreatbox=\hbox{\raise.5ex\hbox{$>$}\llap
     {\lower.5ex\hbox{$\sim$}}}\ht1=\grdimen\dp1=0pt
\setbox\simlessbox=\hbox{\raise.5ex\hbox{$<$}\llap
     {\lower.5ex\hbox{$\sim$}}}\ht2=\grdimen\dp2=0pt
\setbox\simpropbox=\hbox{\raise.5ex\hbox{$\propto$}\llap
     {\lower.5ex\hbox{$\sim$}}}\ht2=\grdimen\dp2=0pt
\def\ga{\mathrel{\copy\simgreatbox}}
\def\la{\mathrel{\copy\simlessbox}}
\def\simprop{\mathrel{\copy\simpropbox}}

\title[Jet models for black-hole binaries]{Jet models for black-hole binaries in the hard spectral state}

\author[A. A. Zdziarski, {\L}.\ Stawarz, P. Pjanka and M. Sikora]
{Andrzej A. Zdziarski,$^1$ {\L}ukasz Stawarz,$^{2,3}$ Patryk Pjanka$^4$ and Marek Sikora$^1$\\
$^1$Centrum Astronomiczne im.\ M. Kopernika, Bartycka 18, PL-00-716 Warszawa, Poland\\
$^2$Institute of Space and Astronautical Science JAXA, 3-1-1 Yoshinodai, Chuo-ku, Sagamihara, Kanagawa 252-5210, Japan\\
$^3$Astronomical Observatory, Jagiellonian University, Orla 171, PL-30-244 Krak{\'o}w, Poland\\
$^4$Obserwatorium Astronomiczne Uniwersytetu Warszawskiego, Al. Ujazdowskie 4, PL-00-478 Warszawa, Poland
}

\date{Accepted 2014 March 2.  Received 2014 March 2; in original form 2013 July 4}

\pagerange{\pageref{firstpage}--\pageref{lastpage}}
\pubyear{2013}

\begin{document}

\maketitle

\label{firstpage}

\begin{abstract}
This is part one of our study of models of jets with distributed electron acceleration. We present here our assumptions, basic equations, and their solutions for the steady-state electron distribution. We assume the shape of the rate of electron acceleration and the dependencies of its normalization and the magnetic field strength on the height along the jet. Our focus is on the hard spectral state of black-hole binaries, for which we take into account that their typical radio spectra are flat. This appears to require a constant dissipation rate per unit logarithmic length and conservation of the magnetic energy flux. Our electron kinetic equation includes adiabatic and radiative losses and advection, and our photon radiative transfer equation includes synchrotron absorption and emission and Compton emission. Apart from the self-Compton process, we take into account Compton scattering of stellar and accretion photons and absorption of very-high energy gamma-rays by pair production on soft photons. We present a general solution of the kinetic equation with advection and radiative and adiabatic losses and an analytic solution in the case of dominant synchrotron losses in conical jets. In the following paper, we present detailed spectra resulting from our equations as applied to Cyg X-1.
\end{abstract}

\section{Introduction}
\label{intro}

Radio-emitting jets are common in accreting black-hole binaries. In their hard spectral state (see, e.g., \citealt*{dgk07}), the jet is approximately steady and characterized by a flat, $F(E)\simprop E^0$, radio spectrum (e.g., \citealt{fender00}). This property is well explained by the particle flux of the radio-emitting electrons (non-thermal with a power-law distribution) being maintained along the jet together with conservation of magnetic energy flux \citep{bk79}. Given the presence of substantial adiabatic and radiative losses, the observation of a flat radio spectrum appears to require a distributed dissipation along the jet.

In accreting binaries with a high-mass donor (e.g., Cyg X-1 and Cyg X-3), the jet is also irradiated by a strong flux of blackbody photons from the donor. The relativistic electrons Compton upscatter those photons at a rate proportional to the product of the electron number and the irradiating flux, which is approximately constant per unit length up to the height of the order of the binary separation. Observationally, hard state jets extend, on one hand, to radii much above the binary separation (e.g., \citealt{stirling01}). On the other hand, the radio-emitting relativistic electrons are present in the jet at least down to the height of the order of the binary separation, as implied by the strong orbital modulation of the radio emission in the case of Cyg X-1 \citep{zdz12}. Thus, hard-state jets (unless they are highly relativistic) can effectively emit Compton upscattered blackbody radiation (hereafter BBC) over very long lengths. Other optically-thin radiative processes, namely synchrotron, synchrotron self-Compton (SSC), and upscattering of photons emitted by the accretion flow (hereafter XC) may dominate the jet emission close to the jet base. Furthermore, jet electrons lose energy adiabatically (Appendix \ref{adiabatic}).

There have been a large number of models proposed to account for global emission properties of the jets in AGN and black hole binaries. First, (i) the radiating part of a jet is approximated as one-zone, even if the shape of the jet and the dependence of the magnetic field on height are sometimes taken into account, e.g., by \citet*{moderski03} and \citet{gt09}. Those models have been applied mostly to blazars. There have been a very large number of studies in this category, some recent papers are \citet{dermer09}, \citet{gt09} or \citet{ghisellini10}. Then, another category of models consider jets with extended emission. In one class (ii) of such models, relativistic electrons are assumed to be accelerated exclusively at the jet base, and then their evolution is studied as they move along the jet (e.g., \citealt{hj88,kaiser06,pc09,pc12}). However, reacceleration appears necessary, e.g., \citet{malzac13}. Then, another class (iii) of models assumes that the electron distribution is maintained along the jet by some unspecified process (e.g., \citealt*{bk79,fb95,heinz06,brp06,zls12}, hereafter ZLS12; \citealt*{mzc13}). In some of these studies, the effect of radiative cooling on the distribution is taken into account by including a cooling break in the electron distribution, at the Lorentz factor at which the rates of radiative and adiabatic losses are equal. The importance of cooling breaks in jets of black-hole binaries was pointed out by, e.g., \citet{hs03} and \citet{heinz04}. The presence of the cooling break is predicted in jets in which electron acceleration occurs in localized regions, which is the case, e.g., in the internal shock model (e.g., \citealt{malzac13}). In many models, however, several simplifying assumptions, e.g., the electron cooling break constant being along the jet, are introduced (e.g., ZLS12). Yet another class (iv) of models assumes the rate of electron acceleration as a function of the height and Lorentz factor, and solves for the electron steady state distribution (e.g., \citealt*{kab08,vila12,reynoso12,pc13}). Similarly to the models (ii), the equation governing that evolution includes radiative cooling together with advection along the jet. The present work belongs to the class (iv). Different than the above studies (iv), we take into account the BBC and XC cooling and emission. 

Generally, the rate of radiative losses strongly varies along the jet, and in high-mass binaries, this dependence is further complicated by the presence of the BBC losses. The relative importance of radiative cooling decreases with the jet height, which effect increases the number of relativistic electrons at high-energies with respect to that in models without a cooling break or with a uniform electron break Lorentz factor. This effect can strongly increase the BBC flux above the MeV energies. 

Here, we present a jet model addressing the above issues in detail. We assume the rate of the electron acceleration to be a power-law function of the electron energy, with both low and high-energy cutoffs. We also anticipate power-law scaling of the dissipation rate and of the magnetic field strength with the jet height. In most of the calculations, we make a simplifying assumption that the jet above certain height is conical and has a constant velocity. We formulate a kinetic equation for the electron distribution at each height taking into account the adiabatic and radiative losses and advection, and a radiative transfer equation taking into account the synchrotron and Compton processes. For both electron distribution and emission spectra, we take into account the effect of synchrotron self-absorption. We solve the resulting equations self-consistently taking into account the SSC, BBC and XC processes and photon-photon pair absorption of \g-ray photons. In the case of adiabatic, optically-thin synchrotron and Thomson losses and advection, we find the analytical solution to the electron kinetic equation. We also introduce some approximate analytical solutions in Appendices \ref{Thomson}--\ref{compton_loss}, and present a method of calculating the opacity to photon-photon pair production in the field of a star in Appendix \ref{taupair}. We assume no efficient acceleration of ions, which process would lead to hadronic radiative processes, taken into account by, e.g., \citet{vila12} and \citet{reynoso12}. We also neglect bremsstrahlung, which is usually inefficient in jets (e.g., \citealt{vila12}). 

Our model is intended to account for the jet contributions to broad-band spectra of accreting sources, from radio to high and very high energy \g-rays. Radio emission is common in accreting X-ray binaries, but so far the only firm case of high-energy \g-ray emission detected from such systems is that of Cyg X-3, which has been detected by \fermi\/ Large Area Telescope (LAT) \citep{fermi} and by \agile\/ \citep{agile}. This emission has been well explained by Compton upscattering of the donor blackbody photons by relativistic electrons in a localized region of the jet \citep{dch10b,z12}. This appears to be the only proposed explanation of the \g-ray emission that accounts for the observed \citep{fermi} strong orbital modulation of the \g-rays. It demonstrates that Compton upscattering of stellar blackbody photons by a jet in a high-mass X-ray binary is an observable process. In the soft state of Cyg X-3, electron acceleration in the inner parts of the jet occurs intermittently. On the other hand, radio data for high-mass black-hole binaries in hard states reveal persistent acceleration along the entire jet body. Then, not only a detection but also upper limits on the $\gamma$-ray emission can constrain relativistic electron spectra in hard-state jets.

The second case of reported high and very high \g-ray emission from an accreting binary is that of Cyg X-1. A single short flare of the source has been reported in the TeV range \citep{magic}. Recently, upper limits at $>30$ MeV and a tentative detection of a steady 0.1--10 GeV emission component in the hard state of Cyg X-1 have been found in the most recent accumulation of the \fermi\/ LAT data \citep{mzc13}. No emission has been found in the soft state. The presence of \g-ray emission in the hard-state has been independently confirmed by \citet{bodaghee13}, who reported 21 (low-significance) detections of Cyg X-1 by \fermi\/ LAT on a daily time scale. Out of those, 15 were in the hard state, five in an intermediate state, and only one event was in the soft state, in spite of a substantial exposure in that state. It is then highly unlikely that the hard/intermediate state detections were statistical fluctuations. This confirms the results of \citet{mzc13}. 

\section{The model setup}
\label{def}

\begin{table}
\centering
\caption{The main symbols used in this work.
}
\begin{tabular}{cc}
\hline
Symbol & Meaning \\
\hline
$\beta_{\rm j}$ & the jet bulk velocity in units of $c$\\
$\Gamma_{\rm j}$ & the jet bulk Lorentz factor\\
$z$ & height along the jet\\
$r_{\rm j}(z)$ & the jet radius\\
$z_{\rm m}$ & the onset of dissipation\\
$z_{\rm M}$ & the end of dissipation\\
$\xi\equiv z/z_{\rm m}$ & dimensionless height\\
$x$ & distance from the jet axis perpendicular to $z$\\
$r_*$ & the stellar radius\\
$T_*$ & the stellar temperature\\
$\Theta$ & the stellar temperature in the jet frame in unit of $m_{\rm e}c^2$\\
$a$ & the separation between the stellar components\\
$i$ & the system inclination\\
$r$ & the distance from a jet point to the donor centre\\
$\phi_{\rm b}$ & the orbital phase\\
$D$ & distance to the system\\
$E$ & observed photon energy\\
${\cal D}_{\rm j,cj}$ & the jet or counterjet Doppler factor\\
${\cal D}_*$ & the Doppler factor of stellar radiation\\
${\cal D}_{\rm d}$ & the Doppler factor of the accretion radiation\\
$\gamma$ & the Lorentz factor of either a jet electron or produced e$^\pm$\\
$\beta$ & the $\beta$ factor corresponding to $\gamma$\\
$\gamma_{\rm m}$ & the minimum injected $\gamma$\\
$\gamma_{\rm M}$ & the injected $\gamma$ corresponding to the high-energy cutoff\\
$Q(\gamma,\xi)$ & the injection rate per unit volume in radiating region\\
$q(\xi)$ & the spatial dependence of $Q$\\
$Q_0$ & the normalization of $Q(\gamma,\xi)$\\
$\tilde{Q}(\gamma,\xi)$ & dimensionless injection rate per unit jet length\\
$\eta_{\rm acc}\leq 1$ & scaling factor for the acceleration rate\\
$p$ & the index of the injected electrons\\
$s$ & the local index of the steady-state electrons\\
$\dot \gamma<0$ & the time derivative of $\gamma$ in the jet frame\\
$\dot\gamma_i$ & components of $\dot\gamma$; $i=$ ad, S, SSC, BBC, XC\\
$\gamma_\xi$ & ${\rm d}\gamma/{\rm d}\xi=\dot\gamma {\rm d}t/{\rm d}\xi$, the rate of change of $\gamma$ with $\xi$\\
$N(\gamma,\xi)$ & the steady-state electron distribution per unit volume\\
$\tilde{N}(\gamma,\xi)$ & dimensionless electron distribution per unit jet length\\
$g_{\rm cut}(\gamma,\gamma_{\rm M})$ & the shape of the cutoff of $Q(\gamma,\xi)$\\
$f_{\rm cut}(\gamma,\gamma_{\rm M})$ & the shape of the cutoff of $N(\gamma,\xi)$\\
$\gamma_0$ & the minimum $\gamma$ to which $N(\gamma,\xi)$ electrons are cooled\\
$\gamma_{\rm b}(\xi)$ & the cooling break\\
$B_0$ & the magnetic field strength at $z_{\rm m}$\\
$B_{\rm cr}$ & the critical magnetic field, $={2\upi m_{\rm e}^2 c^3/(e h)}$\\
$\epsilon$ & jet frame dimensionless photon energy\\
$\epsilon_0$ & dimensionless seed photon (e.g., blackbody) energy\\
$n_0$ & seed photon density/distribution\\
$\theta$ & Compton scattering angle in the jet frame\\
$\chi$ & $1-\cos\theta$\\
$F(E)$ & energy flux ${\rm d}F/{\rm d}E$ ($F$ energy unit = photon $E$ unit)\\
$\Theta_{\rm b}$ & the observed brightness temperature, $\Theta_{\rm b}\equiv k T_{\rm b}/(m_{\rm e}c^2)$\\
$\alpha$ & energy spectral index, $F(E)\propto E^{-\alpha}$\\
$F_k(E)$ & flux from a process; $k=$ S, SSC, BBC, XC\\
$F_{\rm X}$ & flux from the accretion flow\\
$\beta_{\rm eq}$ & the plasma $\beta$ parameter\\
$\sigma_{\rm eq}$ & the magnetization parameter\\
$w$ & enthalpy\\
$P_k$ & jet power; $k=$ inj, ad, S, SSC, BBC, XC, e,  i, $B$\\
$j_k$ & emissivity (per unit solid angle); $k=$ S, SSC, BBC, XC\\
$\alpha_{\rm S}$ & synchrotron absorption coefficient\\
$\tau_{\rm S}$ & synchrotron optical depth\\
$E_{\rm t0}$ & the observed turnover energy\\
$\epsilon_{\rm t0}$ & jet frame dimensionless turnover energy at $z_{\rm m}$\\
$\epsilon_{\rm t}$ & jet frame dimensionless turnover energy at $z$\\
$\gamma_{\rm t}$ & $\gamma$ corresponding to $\epsilon_{\rm t}$\\
$\tau_{\gamma\gamma}$ & pair absorption optical depth\\
\hline
\end{tabular}
\label{symbols}
\end{table}

ZLS12 have developed a global jet emission model applied to Cyg X-1. Here, we extend this work by considering additional effects and processes, listed below: (i) Compton upscattering of stellar photons (BBC) by the jet. As discussed in Section \ref{intro}, upscattering of blackbody photons is important in high-mass X-ray binaries. (ii) Compton upscattering of accretion-flow photons (XC). Only some general energetic constraints were given in ZLS12 for these two effects. (iii) The dependence of the shape of the electron distribution on the height along the jet, $z$, which was neglected in ZLS12. Also, the SSC process was considered by ZLS12 in a one-zone approximation only. (iv) Advection of electrons along the jet, which was neglected in ZLS12. The effects (i--ii) have been included in jet models of Cyg X-1 of \citet{mzc13}. The effects (ii--iv) may be important in both low and high-mass X-ray binaries. 

Table \ref{symbols} lists the main symbols used in this work. In particular, we define the differential energy flux, $F(E)\equiv {\rm d}F/{\rm d}E$, as having the energy unit in $F$ the same as the unit of $E$. As the two energy units cancel each other, the unit of $F(E)$ is cm$^{-2}$ s$^{-1}$. This convention, adopted hereafter, simplifies substantially all the radiative formulae introduced in the paper. The integrated energy flux has then the unit of energy (either erg or keV) times cm$^{-2}$ s$^{-1}$. We express the height along the jet in units of $z_{\rm m}$, $\xi\equiv z/z_{\rm m}$, where $z_{\rm m}$ is at the height of the onset of dissipation and emission. 

We then specify the coordinate system, which is needed to account for the anisotropy effects in the BBC, XC and pair-absorption processes in binaries. We assume the jet and counterjet are perpendicular to the binary plane, i.e., along the $z$ axis, the orbit to be circular, the superior conjunction (compact object behind the star) corresponds to the orbital phase (= polar angle) of $\phi_{\rm b}=0$, and the $x$ axis points away from the star to the direction of the compact object at the superior conjunction. Then, the unit vectors pointing towards the observer, from the stellar centre to the compact object, along the jet and counterjet, from the stellar centre to the emission location along the jet at the height $\pm z$, and from a point on the accretion disc distant from the centre by $r_{\rm d}$ and with the azimuth $\phi$ are, respectively,
\begin{eqnarray}
\lefteqn{\vec{e}_{\rm obs}=(-\sin i, 0, \cos i),\nonumber}\\
\lefteqn{\vec{e}_{\rm c}=(\cos\phi_{\rm b},\sin\phi_{\rm b}, 0),\nonumber}\\
\lefteqn{\vec{e}_{\rm j}=(0, 0, 1),\quad \vec{e}_{\rm cj}=(0, 0, -1),\nonumber}\\
\lefteqn{\vec{e}_*=\left({a\over r}\cos\phi_{\rm b},{a\over r}\sin\phi_{\rm b}, \pm {\xi z_{\rm m}\over r}\right),}\\
\lefteqn{\vec{e}_{\rm d}=\left({r_{\rm d}\over r_{\rm dj}}\cos\phi,{r_{\rm d}\over r_{\rm dj}}\sin\phi, \pm {\xi z_{\rm m}\over r_{\rm dj}}\right),\nonumber}\\
\lefteqn{r=\left[a^2+(\xi z_{\rm m})^2\right]^{1/2},\nonumber}\\
\lefteqn{r_{\rm dj}=\left[r_{\rm d}^2+(\xi z_{\rm m})^2\right]^{1/2},\nonumber}
\end{eqnarray}
where the $+$ and $-$ sign is for the jet and counterjet, respectively, and $r$ is the distance from the stellar centre to the emission location. 

We define a dimensionless photon energy in the jet/counterjet frame and the jet and counterjet Doppler factors (for perpendicular jets),
\begin{equation}
\epsilon={E\over {\cal D}_{\rm j,cj}(\xi) m_{\rm e} c^2},\quad {\cal D}_{\rm j,cj}(\xi)={1\over \Gamma_{\rm j}(\xi)(1\mp\beta_{\rm j}(\xi)\cos i)},
\label{rel}
\end{equation}
where $E$ is the observed dimensional photon energy, the $-$ and $+$ sign is for the jet and counterjet, respectively, $\beta_{\rm j} c$ is the jet bulk velocity, and $\Gamma_{\rm j}$ is the corresponding Lorentz factor. We assume the entire jet motion is along its axis, which is a sufficient approximation for narrow jets. We also define a Doppler factor of the stellar radiation (at $E_0$) seen in the jet frame (at $\epsilon_0$),
\begin{equation}
{\cal D}_*={1\over \Gamma_{\rm j}(\xi)\left\{1-\beta_{\rm j}(\xi)/[1+(a/\xi z_{\rm m})^2]^{1/2}\right\}},\quad \epsilon_0={E_0\over {\cal D}_*(\xi) m_{\rm e} c^2},
\label{dstar}
\end{equation}
which ${\cal D}_*$ approximates the star as a point source, and is the same for the jet and counterjet in the perpendicular case assumed here. Analogously, the jet-frame Doppler factor of emission from a point distant by $r_{\rm d}$ from the jet axis on the plane at $z=0$ (e.g., of an accretion disc) is,
\begin{equation}
{\cal D}_{\rm d}={1\over \Gamma_{\rm j}(z)\left\{1-\beta_{\rm j}(z)/\left[1+(r_{\rm d}/z)^2\right]^{1/2}\right\} }.
\label{dx}
\end{equation}
The relationship between photon energy in the jet frame to that in the system frame is the same as above.

We consider jets of an arbitrary shape, given by the dependence of the radius on height, $r_{\rm j}(\xi)$, the magnetic field dependence, $B(\xi)$, the normalization of the dissipation rate per unit volume, $q(\xi)$, and the velocity, $\beta_{\rm j}(\xi)$. However, we analyze in detail only conical jets with conserved magnetic energy flux and the dissipation rate (at a given $\gamma$) constant per $\ln \xi$, and constant velocity,
\begin{equation}
r_{\rm j}(\xi)=z_{\rm m} \xi\tan\Theta_{\rm j},\quad B(\xi)=B_0\xi^{-1},\quad q(\xi)=Q_0\xi^{-3},\quad \beta_{\rm j}(\xi)=\beta_{\rm j},
\label{conical}
\end{equation}
where $\Theta_{\rm j}$ is the opening angle, $B_0$ is the field strength at the jet base, and $Q_0$ is a constant (see below).

\section{The electron distribution}
\label{distribution}

We assume continuous energy dissipation along the jet. We also assume that the outflow is composed of two types of regions, one characterized by efficient dissipation of the jet energy and conversion of it into electron acceleration, and the other dominated by electron energy loss processes (as, e.g., in the model of \citealt*{krm98}). In the former regions, which we assume to occupy only a small part of the jet, electrons undergo a diffusive acceleration/escape process limited solely by radiative losses. We discuss possible forms of the acceleration. Then we consider in detail the emission regions, taking into account injection of the accelerated electrons, electron energy losses and advection of electrons along the jet. The jet is launched at some height, $z_0$, and it is then accelerated. The dissipation is assumed to begin at some height $z_{\rm m}>z_0$, and to end at $z_{\rm M}\gg z_{\rm m}$. Above $z_{\rm M}$, the jet continues to propagate, and we follow the electron evolution until the electrons lose all their energies.

\subsection{Acceleration region}
\label{acceleration}

\begin{figure}
\centerline{\includegraphics[width=0.8\columnwidth]{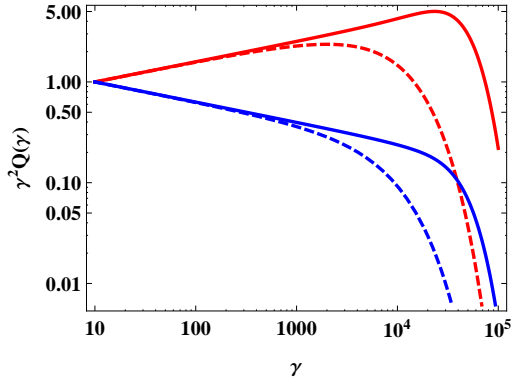}} 
\caption{The distribution of accelerated electrons in the case of quasi-linear stochastic acceleration with $q_{\rm turb}=2$ \citep{sp08} for $p=1.8$ and 2.2 (red and blue solid curves, respectively) and $\gamma_{\rm M}=10^4$, compared with the corresponding e-folded power-law distributions (dashed curves). We see that the e-folded power laws strongly underestimate the actual rates around the cutoff. The normalization is to unity at $\gamma=10$.
} \label{acc}
\end{figure}

Physical mechanisms resulting in the efficient electron acceleration within relativistic outflows are still hardly known, being subjected to the ongoing debate. It could be diffusive shock acceleration (e.g., \citealt{ma12,malzac13}), stochastic acceleration by turbulent magnetic field (e.g., \citealt{sp08,asano14}) or direct electron energization within magnetic reconnection sites (e.g., \citealt*{lyutikov06,giannios09}). Given this lack of knowledge, we assume the distribution of accelerated electrons forms a power law, $Q(\gamma)\propto \gamma^{-p}$, between low and high-energy cutoffs, and we take $p$ to be a free parameter. We note that the non-thermal tails in the particle spectra formed due to diffusive shock acceleration in the case of relativistic outflows are expected to be approximately of a power-law form, with the energy index $p \geq 2$ depending on the shock velocity, magnetic field orientation (with respect to the shock normal), and the turbulence conditions in the vicinity of a shock (see, e.g., \citealt{no04,ss09,ss11}). 

The high-energy cutoff of the distribution, $\gamma_{\rm M}$, is, most likely, related to the acceleration rate balanced by radiative losses within the acceleration region. The acceleration rate in the high-efficiency limit can be parameterized as $\dot \gamma_{\rm acc} m_{\rm e} c=\eta_{\rm acc} e B$, where $m_{\rm e}$ and $e$ are the electron mass and charge, respectively, and $\eta_{\rm acc}$ is an efficiency factor, which has to be $\leq 1$ unless the electric field is $>B$. We then balance this rate by that of synchrotron losses (though we check the effect of inclusion of other losses). This gives $\gamma_{\rm M}$ and the corresponding cutoff synchrotron photon energy,
\begin{equation}
\gamma_{\rm M} \simeq \left(9 \eta_{\rm acc}B_{\rm cr} \over 4\alpha_{\rm f} B\right)^{1/2},\quad \epsilon_{\rm M}\simeq \gamma_{\rm M}^2 {B\over B_{\rm cr}}= {9\eta_{\rm acc} \over 4 \alpha_{\rm f}}\simeq 310 \eta_{\rm acc},
\label{gemax}
\end{equation}
where $\alpha_{\rm f}=2\upi e^2/(h c)$ is the fine-structure constant, $B_{\rm cr}$ is the critical magnetic field, see Table \ref{symbols}, and $h$ is the Planck constant. We note $\epsilon_{\rm M}$ is independent of $B$ \citep*{gfr83,dj96}. For conserved magnetic energy flux in a conical jet, equation (\ref{conical}), $\gamma_{\rm M}\propto \xi^{1/2}$.

We note several caveats to the above relatively standard assumption. Electric field $>B$ seems to occur occasionally in the Crab Nebula, and thus $\eta_{\rm acc}$ can be $> 1$ (see, e.g., \citealt{cerutti12,cerutti13}). On the other hand, the rate of the diffusive (and, in particular, the diffusive shock) acceleration is limited by the Larmor period, $2\upi r_{\rm L}/c$, where $r_{\rm L}=\gamma m_{\rm e} c^2/(e B)$ is the Larmor radius, which constraint corresponds to $\eta_{\rm acc}\leq 1/2\upi$. (This factor of $2\upi$ was included in the definition of $\eta_{\rm acc}$ in ZLS12.) Then, the Larmor radius has to be less than the size of a single acceleration region, $R_{\rm acc}$, which imposes an independent constraint of $\gamma_{\rm M}<e B R_{\rm acc}/(m_{\rm e}c^2)$, and $\epsilon_{\rm M}\propto B^3$. Furthermore, the acceleration process can be purely stochastic in nature, with the acceleration rate depending on the scale distribution of the turbulence energy density, usually parameterized by its power-law index, $q_{\rm turb}$. In the Bohm limit of $q_{\rm turb}=1$, $\gamma_{\rm M}$ is given approximately by equation (\ref{gemax}), but the steady-state spectrum of the accelerated particles is expected to be flat, with $p = 0$ (e.g., \citealt{bs85,sp08}). On the other hand, a range of $p\geq 1$ can be obtained in the 'hard-sphere' limit of $q_{\rm turb}=2$ \citep{schlickeiser84,pp95}, but then the acceleration time scale is approximately constant, which implies $\gamma_{\rm M}\propto B^{-2}$ (though still limited by the overall maximum of equation \ref{gemax}). This corresponds to $\epsilon_{\rm M}\propto B^{-3}$. Finally, we note that the values of $B$ in the acceleration and energy loss regions can be different.

Since we do not know which acceleration process operates in jets, we cannot uniquely specify the shape of the high-energy cutoff in the acceleration rate. For example, in the stochastic case with $q_{\rm turb}=2$, it is given by equation (8) in \citet{sp08}. Fig.\ \ref{acc} shows a comparison of that solution to an e-folded power law for two values of $p$. We see e-folded spectrum strongly underestimates the solution. Thus, we introduce a symbolic function, $g_{\rm cut}(\gamma, \gamma_{\rm M})$, to denote the cutoff in the spectrum of the accelerated particles. However, since we use here a $\delta$-function approximation to the synchrotron emissivity, finding the accurate form of the electron cutoff is out of scope of this paper. 

At low energies, we also expect a cutoff at $\gamma_{\rm m}>1$, likely at $\gg 1$. E.g., a quasi-Maxwellian electron distribution with $\bar\gamma \gg 1$ has been found in particle-in-cell simulations of collisionless shocks \citep{s08,ss11}. In those simulations, an approximate energy equipartition between quasi-thermal electrons and ions is often observed, namely $\bar\gamma \simeq (m_{\rm i}/m_{\rm e}) (\bar \gamma_{\rm i}-1)$, where $\bar \gamma_{\rm i}$ is the average Lorentz factor of the ions and $m_{\rm i}$ is the ion mass. Acceleration then proceeds only from this distribution. Here, we approximate this situation by a sharp low-energy cutoff at $\gamma_{\rm m}\simeq \bar\gamma$. This differs, e.g., from the approach of \citet{gt09}, who assumed a hard injection below $\gamma_{\rm m}$.

Thus, the form of the electron acceleration we use is
\begin{equation}
Q(\gamma,\xi) \simeq \cases{0,&$\gamma<\gamma_{\rm m}$;\cr
q(\xi) \gamma^{-p}g_{\rm cut}(\gamma, \gamma_{\rm M}), &$\gamma\geq \gamma_{\rm m}$.\cr}
\label{Q_inj1}
\end{equation}
When $q(\xi)=Q_0\xi^{-3}$, equation (\ref{conical}), we have a constant injected power per unit $\gamma$ per unit logarithmic interval of the height of a conical jet. Together with the conserved magnetic energy flux, this leads to the spectral index $\alpha = 0$ in the partially self-absorbed segment of the synchrotron spectrum independently of the value of the electron power-law index, $p$ \citep{bk79}, which is approximately satisfied in the hard state of black-hole binaries, in particular in Cyg X-1 \citep{fender00}. On the other hand, \citet{vila12} assumed $q(\xi)=$ const, which corresponds to the power per unit $\ln z$ being $\propto z^3$, i.e., most of it injected around $z_{\rm M}$, which appears to us rather unlikely. 

We note that the condition of $q(\xi)\propto \xi^{-3}$ is somewhat different from the corresponding condition for the integrated power per $\ln\xi$, $\propto \int_{\gamma_{\rm m}}^\infty {\rm d} \gamma\,\gamma Q(\gamma,\xi)$, especially for $p<2$ due to the dependence of $\gamma_{\rm M}\propto \xi^{1/2}$ in conical jets with conserved magnetic energy flux, implying the integrated power is $\propto \xi^{1-p/2}$. In this case, it is possible that $\gamma_{\rm M}$ is limited not by synchrotron losses but by the available injected power, and thus $\gamma_{\rm M}$ could be constant with $\xi$. However, given theoretical uncertainties about the dissipation rate and the acceleration mechanism, we keep below the relation of equation (\ref{gemax}). For $p>2$, the leading dependence is on $\gamma_{\rm m}$, which is likely to be constant with height. In that case, the dependencies of $Q(\gamma,\xi)$ and  $\int_{\gamma_{\rm m}}^\infty {\rm d}\gamma\,\gamma Q(\gamma,\xi)$ on $\xi$ are almost the same.

\subsection{Emission region}
\label{emission}

\subsubsection{Energy loss rates}
\label{loss_rates}

The electron energy loss rate, $\dot \gamma$, in an expanding jet irradiated by a companion star and an accretion flow has in general five components. These are due to the adiabatic, synchrotron, BBC, SSC and XC processes, with the corresponding rates denoted below by $\dot \gamma_{\rm ad}$, $\dot \gamma_{\rm S}$, $\dot \gamma_{\rm BBC}$, $\dot \gamma_{\rm SSC}$, $\dot \gamma_{\rm XC}$, respectively.

The adiabatic loss rate (see also Appendix \ref{adiabatic}) is given by
\begin{equation}
-\dot\gamma_{\rm ad}={2\over 3}{{\rm d}\ln r_{\rm j}(\xi)\over {\rm d}\xi}{{\rm d} \xi\over {\rm d}t}(\gamma-1),\quad {{\rm d} \xi\over {\rm d}t} ={c \Gamma_{\rm j}\beta_{\rm j}(\xi)\over z_{\rm m}},
\label{gdot_ad}
\end{equation}
where $t$ is time in the comoving frame. In a conical jet,
\begin{equation}
-\dot\gamma_{\rm ad}=A_{\rm ad} {\gamma-1\over \xi},\quad A_{\rm ad}={2\Gamma_{\rm j}\beta_{\rm j}(\xi)c\over 3 z_{\rm m}}.
\label{radius}
\end{equation}

The synchrotron loss rate can be written as
\begin{eqnarray}
\lefteqn{
-\dot\gamma_{\rm S}\approx {\sigma_{\rm T} B^2(\gamma^2-1)\over 6\upi m_{\rm e} c} {1-\exp[-\tau_\perp(\gamma,\xi)]\over \tau_\perp(\gamma,\xi)}, \label{gdot_syn}}\\
\lefteqn{
\tau_\perp(\gamma,\xi)\equiv \alpha_{\rm S}(\gamma,\xi)z_{\rm m}\xi\tan\Theta_{\rm j}, 
\label{tau_syn}}
\end{eqnarray}
where $\tau_\perp$ is the radial (perpendicular to the jet axis) self-absorption optical depth, the absorption coefficient, $\alpha_{\rm S}(\gamma, \xi)$, is given by equation (\ref{alphas}) below, and we relate the synchrotron $\epsilon$ and $\gamma$ by the delta-function approximation, $\epsilon=(B/B_{\rm cr})\gamma^2$. The term involving $\tau_\perp$ follows from the equation of radiative transfer and the relation between $\dot\gamma_{\rm S}$ and the corresponding emissivity. The optical depth here should have been averaged over the jet cross section and all directions, but here we use $\tau_{\perp}$ for simplicity. The second term above can also be approximated in a simpler form as $(1+\tau_\perp)^{-1}$. For $B(\xi)=B_0/\xi$, equation (\ref{conical}), we therefore rewrite the synchrotron energy loss rate as
\begin{equation}
-\dot\gamma_{\rm S}\approx {A_{\rm S}(\gamma^2-1)\over \xi^2[1+\tau_\perp(\gamma,\xi)]}, \quad A_{\rm S}\equiv{\sigma_{\rm T} B_0^2\over 6\upi m_{\rm e} c}.
\label{gdot_syn0}
\end{equation}

The loss rate due to Compton scattering of stellar blackbody photons is
\begin{eqnarray}
\lefteqn{
-\dot \gamma_{\rm BBC}=
{A_{\rm BBC}(\xi) (\gamma^2-1)\over 1+\xi^2 z_{\rm m}^2/a^2},\label{gdot_bbc}}\\
\lefteqn{
A_{\rm BBC}(\xi) ={8\upi^5(k T_*)^4 \sigma_{\rm T}f_{\rm KN}(\gamma,\xi)\over 45 m_{\rm e}c^4 h^3}\left[r_*\over a {\cal D}_*(\xi)\right]^2,}
\end{eqnarray}
where $f_{\rm KN}$ is the Klein-Nishina (KN) correction factor ($=1$ in the Thomson limit), and $\sigma_{\rm T}$ is the Thomson cross section. This formulation of BBC losses uses $\dot \gamma$ averaged over all the directions of electrons, assuming the efficient particle isotropization (note that the electrons moving towards the star lose the energy faster than the ones moving in the opposite direction). We note here that $A_{\rm BBC}$, rather than being a constant, includes the dependence of ${\cal D}_*$ on $\xi$. However, this dependence is monotonic, and weak as long as $\beta_{\rm j}$ is not close to 1, 
\begin{equation}
{1\over \Gamma_{\rm j}(\xi){\cal D}_*(\xi)}=1-{\beta_{\rm j}(\xi)\xi z_{\rm m}/ a\over \left(1+\xi^2 z_{\rm m}^2/a^2\right)^{1/2}}\simeq\cases{1,&$\xi\ll a/z_{\rm m}$;\cr
1-\beta_{\rm j}(\xi), &$\xi\gg a/z_{\rm m}$.}
\label{dstar_a}
\end{equation}
Thus, it does not affect dominant functional dependence of $\dot \gamma_{\rm BBC}(\xi)$. 

Also, $A_{\rm BBC}$ includes the factor $f_{\rm KN}$, which depends on $\xi$ and $\gamma$ outside the Thomson regime. We define it as the ratio of $\dot \gamma_{\rm BBC}$ using the KN cross section to that in the Thomson limit. For $\gamma\gg 1$, it can be obtained from integration of the KN loss rate of \citet{jones68}, 
\begin{eqnarray}
\lefteqn{
-\dot\gamma_{\rm KN}=3\sigma_{\rm T}c\gamma \int_0^\infty {{\rm d}\epsilon_0\over q^2}\,  n_0(\epsilon_0,\xi) \times \nonumber}\\
\lefteqn{ 
\left[
\left({q\over 2} +6 +{6\over q}\right)\ln (1+q) -{11 q^3/12
+2q^2 +q\over (1+q)^2} -6 +2{\rm Li}_2 (-q)\right],
\label{gdotkn}}
\end{eqnarray}
where $q\equiv 4\gamma\epsilon_0$, ${\rm Li}_2$ is the dilogarithm, and $n_0$ is the density of seed photons at $\xi$ and $\epsilon_0$ (in the jet frame). An approximate form of $\dot\gamma_{\rm KN}$ for low $\gamma$ can be obtained by multiplying it by $1-\gamma^{-2}$, analogously to the case of the Thomson rate.

\begin{figure}
\centerline{\includegraphics[width=0.8\columnwidth]{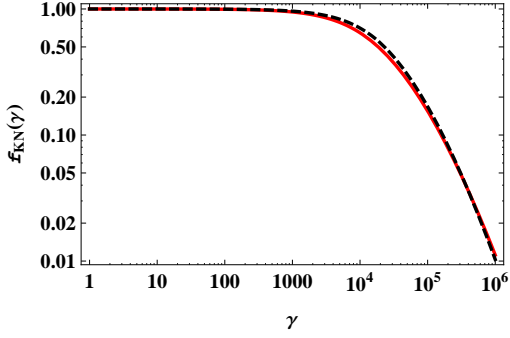}} 
\caption{The KN correction factor for blackbody photons at $T=1.4\times 10^4$ K (solid curve), compared to the KN correction factor for monoenergetic photons at the energy of $2.7 k T$ (i.e., the average energy of blackbody photons; dashed curve).
} \label{fkn}
\end{figure}

In the case of stellar emission, we approximate the star as a point source, in which case the photon beam arriving at the jet is monodirectional, in which case $n_0 =\dot n_0/c$. Stellar photons in the jet frame have a diluted blackbody distribution, 
\begin{equation}
n_{\rm bb}(\epsilon_0,\xi)=\left(m_{\rm e}c\over h \right)^3 \left(r_*\over r\right)^2 {2\upi \epsilon_0^2 {\cal D}_*^2\over \exp(\epsilon_0 /\Theta)-1}\,\, {\rm cm}^{-3},
\label{bb}
\end{equation}
where $\Theta \equiv k T_*/(m_{\rm e}c^2 {\cal D}_*)$. The geometric dilution factor with respect to the blackbody density is given by $\Delta\Omega/4\upi=[r_*/(2 r)]^2$, where $\Delta\Omega$ is the solid angle subtended by the star; it can also be obtained from the relation between the flux at $r$ and the specific intensity from a uniformly-emitting sphere. In addition, the substitution (\ref{dstar}) has been applied. Note that $r$, $\Theta$ and ${\cal D}_*$ depend on $\xi$. In Fig.\ \ref{fkn}, we show $f_{\rm KN}(\gamma)$ for the stellar temperature of the donor in Cyg X-1 at $\xi\gg a/z_{\rm m}$ (where the Compton cooling may dominate), in which case ${\cal D}_*=2$, and the photons have the temperature of $T=T_*/2$ in the jet frame. We see that this $f_{\rm KN}$ is very similar (within $\pm 10$ per cent in the considered range of $\gamma$) to that for monoenergetic photons at the average blackbody energy of $2.7 k T$. (Another approximation to $f_{\rm KN}$ for blackbody photons is provided by \citealt{bk09}.) We also see in Fig.\ \ref{fkn} that the Thomson approximation for the above parameters is satisfactory for $\gamma\la 10^4$. 

Then, we use equation (\ref{gdotkn}) to calculate $\dot \gamma_{\rm SSC}$. In this case, the seed photon density is that of synchrotron photons, $n_{\rm S}$, as given by equation (\ref{sdensity}) below. 

To calculate $\dot \gamma_{\rm XC}$, we need to consider the emission of an accretion flow. That flow usually consists of an optically-thick accretion disc and a hot flow. The local emission of an accretion disc corresponds to,
\begin{equation}
{{\rm d}n_{\rm d}(\epsilon_{\rm d}, r_{\rm d})\over {\rm d}\Omega}=\left(m_{\rm e} c\over  h\right)^3 {1\over f_{\rm c}^4}{2\epsilon_{\rm d}^2 \over \exp\left[\epsilon_{\rm d}/\Theta_{\rm d}(r_{\rm d})\right]-1}\, {\rm cm}^{-3}{\rm sr}^{-1},
\label{disc_bb}
\end{equation}
where $\epsilon_{\rm d}$ is the dimensionless photon energy in the rest frame, $\Theta_{\rm d}\equiv k T_{\rm d}/m_{\rm e}c^2$ is the dimensionless local colour temperature and $f_{\rm c}$ is the colour correction, which is typically $\simeq 1.7$--2 \citep{st95}. The disc extends from $r_{\rm in}$ to $r_{\rm out}$, and the temperature depends on the radius, $r_{\rm d}$,
\begin{equation}
T_{\rm d}(r_{\rm d})=T_{\rm in}\left(r_{\rm d}\over r_{\rm in}\right)^{-3/4},
\label{discT}
\end{equation}
where $T_{\rm in}$ is the maximum disc temperature. Here, we have neglected the inner boundary condition, which can be trivially taken into account for $R_{\rm in}$ close to the innermost stable orbit. The observed flux is given by
\begin{equation}
F_{\rm d}(E)={2\upi c E\cos i\over m_{\rm e}c^2 D^2}\int_{r_{\rm in}}^{r_{\rm out}} {\rm d}r_{\rm d}\,r_{\rm d} {{\rm d}n_{\rm d}(E/m_{\rm e}c^2,r_{\rm d})\over {\rm d}\Omega},
\label{discF}
\end{equation}
whereas the disc blackbody photon density in the jet frame is
\begin{equation}
n_{\rm d}(\epsilon_0, z)=2\upi \int_{\mu_{\rm out}}^{\mu_{\rm in}}{\rm d}\mu_{\rm d} {{\rm d}n_{\rm d}(\epsilon_0{\cal D}_{\rm d}, r_{\rm d})\over {\rm d}\Omega},\quad r_{\rm d}=z\left(\mu_{\rm d}^{-2}-1\right)^{1/2},
\label{disc_density}
\end{equation}
where $\mu_{\rm in}$ and $\mu_{\rm out}$ correspond to $r_{\rm in}$ and $r_{\rm out}$, respectively. We can then substitute equation (\ref{disc_density}) in equation (\ref{gdotkn}). Note that the contribution to the jet-frame density from outer disc regions is, on one hand, reduced due to the small solid angle subtended per unit $r_{\rm d}$ [as compared to the density of photons in the observer's direction, for which equation (\ref{discF}) would have a factor of $\mu^{-3}$ if written as an integral over $\mu$], and, on the other hand, enhanced due to relativistic beaming. The disc emission in the jet frame has also been treated in a number of papers, e.g., \citet{gt09} and \citet{dermer09}. 

Then we describe the emission of the hot flow by an e-folded power law with a Rayleigh-Jeans form at low energies, which approximates thermal Comptonization of the inner disc emission,
\begin{equation}
F_{\rm X}(E) = {K_{\rm X} \left(E_{\rm b}\over 1\,{\rm keV}\right)^{-\alpha_{\rm X}} \exp\left(-E\over E_{\rm c}\right)\over 
\left(E\over E_{\rm b}\right)^{-2} + \left(E\over E_{\rm b}\right)^{\alpha_{\rm X}}},
\label{flux_X}
\end{equation}
where $\alpha_{\rm X}$ is the energy index, $E_{\rm b}\sim k T_{\rm in}$, $E_{\rm c}$ is the e-folding energy, and $K_{\rm X}$ is the normalization. We assume the size of the hot flow is $\ll z_{\rm m}$, in which case its emission forms a monodirectional beam incident on the jet from below, and the Doppler factor of equation (\ref{dx}) becomes ${\cal D}_{\rm X}= {\cal D}_{\rm d}(\mu_{\rm d}=0)=\Gamma_{\rm j}(1+\beta_{\rm j})$. Then, the seed photon density in the jet frame equals
\begin{equation}
n_{\rm X}(\epsilon_0,\xi)={D^2\over c z^2 {\cal D}_{\rm X}\epsilon_0} F_{\rm X}[{\cal D}_{\rm X}\epsilon_0 m_{\rm e}c^2].
\label{density_X}
\end{equation}
We note that \citet{gt09} calculate instead the total energy density of the X-ray photons in the jet frame. That energy density can be used for calculating $\dot\gamma$ only in the Thomson limit, whereas the X-rays are usually deep in the Klein-Nishina limit for the relativistic electrons responsible for high-energy jet emission. 

\subsubsection{The electron kinetic equation and its solutions}
\label{kinetic}

\begin{figure}
\centerline{\includegraphics[width=0.8\columnwidth]{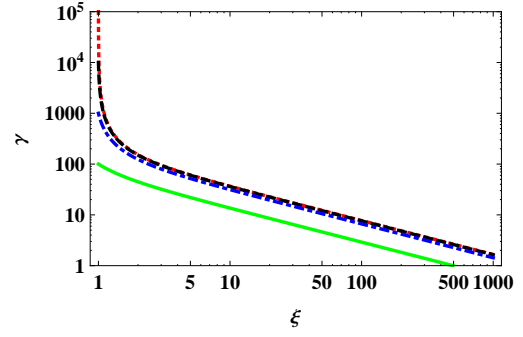}} 
\caption{Exemplary evolutional tracks of electron Lorentz factors along the jet,, $\gamma(\xi)$, for ${A_{\rm S}/ A_{\rm ad}}=0.015$ and the initial values at $\xi=1$ of $10^2$, $10^3$, $10^4$, $10^5$, shown by the green solid, blue dot-dashed, black dashed and red dotted curves, respectively. The evolution follows equation (\ref{gamma_xi}). Its initial bent trajectory is caused by the synchrotron losses, and the straight lines correspond to adiabatic losses only. 
} \label{gamma_vs_xi}
\end{figure}

We assume that the steady-state electron distribution per unit volume in the emission region, $N$, follows the continuity equation in both spatial and energy dimensions with the injection of electrons at the rate of $Q(\gamma,z)$ distributed within the entire volume of a jet. We first consider a general case of an axially symmetric jet with arbitrary profiles $r_{\rm j}(\xi)$, $\beta_{\rm j}(\xi)$, and $\dot\gamma(\gamma,\xi)$, 
\begin{equation}
{1\over z_{\rm m} r_{\rm j}(\xi)^2} {\partial\over \partial \xi}\left[\Gamma_{\rm j}\beta_{\rm j}(\xi)c r_{\rm j}(\xi)^2 N(\gamma,\xi)\right]+{\partial\over \partial \gamma}\left[\dot \gamma(\gamma,\xi) N(\gamma,\xi)\right]=Q(\gamma,\xi).  
\label{ndot}
\end{equation}
The first term in equation (\ref{ndot}), describing spatial advection, corresponds to the divergence in a spherical geometry, $\nabla \cdot N \mathbf{v}$. The second term describes advection in the energy space due to energy losses \citep{kardashev62}. The continuity equation (\ref{ndot}) neglects spatial and momentum diffusion of the radiating particles, but these are relatively minor effects within the emission-dominated jet regions. We note that equation (\ref{ndot}) can also be used for acceleration at the jet base only, in which case $Q$ is given by a $\delta$-function in $\xi$.

To solve this equation, it is convenient to introduce dimensionless distributions integrated over the jet cross section,
\begin{equation}
\tilde{N}(\gamma,\xi)\equiv \upi\Gamma_{\rm j}\beta_{\rm j}(\xi)z_{\rm m} r_{\rm j}(\xi)^2 N(\gamma,\xi),\quad \tilde{Q}(\gamma,\xi)\equiv {\upi r_{\rm j}(\xi)^2 z_{\rm m}^2\over c} Q(\gamma,\xi),
\label{int_dis}
\end{equation}
and the dimensionless loss rate per unit $\xi$,
\begin{equation}
\gamma_\xi\equiv \dot \gamma{{\rm d} t\over {\rm d}\xi}=
{\dot \gamma z_{\rm m}\over c \Gamma_{\rm j}\beta_{\rm j}(\xi)}.
\label{gdot_xi}
\end{equation}
Then, equation (\ref{ndot}) can be rewritten as
\begin{equation}
{\partial\over \partial \xi}\tilde{N}(\gamma,\xi)+{\partial\over \partial \gamma}\left[\gamma_\xi \tilde{N}(\gamma,\xi)\right] =\tilde{Q}(\gamma,\xi),
\label{ndotc}
\end{equation}
which has the same form as corresponding equations in \citet{moderski00,moderski03} and \citet{stawarz08}.

A non-relativistic version of equation (\ref{ndotc}) was considered by \citet{kab08}. (However, it appears that the injection rate in their equation 17 should not be multiplied by the velocity.) Some versions of equation (\ref{ndotc}) have also been used by \citet{vila12} and \citet{reynoso12} to describe the evolution of the electron distribution per unit volume in a conical jet. They, however, appear incorrect, as then equation (\ref{ndot}) should be used.

In solving equation (\ref{ndotc}), we use the fact that an electron with a Lorentz factor $\gamma$ at a height $\xi$ has a unique dependence at other heights, $\xi'$, given by $\gamma'(\gamma, \xi;\xi')$. Here, the first pair of arguments denote the reference values, and $\xi'$ is the main argument of $\gamma'$. It is given by the solution of the differential equation
\begin{equation}
{{\rm d} \gamma'\over {\rm d}\xi'}= \gamma_\xi(\gamma',\xi')
\label{dot_gamma0}
\end{equation}
with the boundary condition of $\gamma'=\gamma$ at $\xi'=\xi$. We note that there may be a minimum value of $\xi' \geq 1$ for which the above solution exists, which we denote as $\xi_{\rm m}(\gamma,\xi)$. The solution of equation (\ref{ndotc}) satisfying the null boundary condition at $\xi=1$ can then be obtained using the method of characteristics (cf.\ \citealt{stawarz08}) as
\begin{eqnarray}
\lefteqn{
\tilde{N}(\gamma, \xi) = \exp\left\{-\int^\xi_{\xi_{\rm m}(\gamma,\xi)}\!\!\! {\rm d}\xi' {\partial\gamma_\xi[\gamma'(\gamma,\xi;\xi'),\xi']\over \partial \gamma'} \right\}
\times \label{general} }\\
\lefteqn{\quad
\int^\xi_{\xi_{\rm m}(\gamma,\xi)}\!\!\!\!\! {\rm d}\xi' \tilde{Q}[\gamma'(\gamma,\xi; \xi'), \xi']  \exp\left\{\!\! \int^{\xi'}_{\xi_{\rm m}(\gamma,\xi)}\!\!\!\!\! {\rm d}\xi'' {\partial\gamma_\xi[\gamma''(\gamma,\xi;\xi''),\xi'']\over \partial \gamma''} \right\}  .\nonumber}
\end{eqnarray}
The derivative of $\gamma_\xi$ can be calculated analytically for adiabatic and synchrotron losses as well as for the Klein-Nishina losses on blackbody radiation approximated as mono-energetic radiation (see below). Then, the terms in $\partial \gamma_\xi/\partial\gamma'$ corresponding to the adiabatic and synchrotron cooling may be integrated analytically, but the Klein-Nishina term needs to be integrated numerically. Thus, a numerical solution of equation (\ref{general}) involves single and double numerical integrations over $\gamma'$ also calculated numerically as the solution of the ordinary differential equation (\ref{dot_gamma0}). In addition, $\xi_{\rm m}$ needs to be calculated numerically, but only once for given $\gamma$ and $\xi$. If Klein-Nishina cooling involves more complex soft photon spectra, e.g., of the accretion disc, there will be one additional internal numerical integration of the Klein-Nishina cooling rate (\ref{gdotkn}) over that spectrum, i.e., the solution will involve a triple numerical integral. Alternatively, equation (\ref{ndotc}) can be solved numerically.

However, the solution can be greatly simplified in the case of adiabatic losses and radiative losses being only synchrotron and Thomson with arbitrary dependencies on $\xi$. For that, we use the results of \citet{stawarz08} given in their equations (A3--A8). From equation (\ref{gdot_ad}), $\gamma_\xi$ for adiabatic losses is
\begin{equation}
\gamma_{\rm \xi,ad}(\gamma,\xi)=-{2\over 3}{{\rm d}\ln r_{\rm j}(\xi)\over {\rm d}\xi}(\gamma-1), 
\label{gamma_ad}
\end{equation}
and $\gamma_\xi$ for optically-thin synchrotron and Thomson losses equals
\begin{equation}
\gamma_{\rm \xi, rad}(\gamma,\xi)=- c_1(\xi) U(\xi) (\gamma^2-1), \quad c_1(\xi)\equiv {4\sigma_{\rm T}z_{\rm m}\over 3 m_{\rm e}c^2 \Gamma_{\rm j}\beta_{\rm j}(\xi)},
\label{dot_gamma_xi}
\end{equation}
where $U(\xi)$ is the sum of the radiation and magnetic energy densities. Equation (\ref{dot_gamma0}) can then be solved for $(\gamma'-1)$. Its somewhat simpler solution for $\gamma'\gg 1$ with the boundary condition of $\gamma'=\gamma$ at $\xi'=\xi$ is
\begin{equation} 
\gamma'(\gamma,\xi;\xi')={\gamma [r_{\rm j}(\xi')/r_{\rm j}(\xi)]^{-2/3}\over 1- \gamma \int_{\xi'}^{\xi} {\rm d}\xi'' c_1(\xi'') U(\xi'') [r_{\rm j}(\xi'')/r_{\rm j}(\xi)]^{-2/3}}.
\label{gamma_xi0}
\end{equation}
Then, equation (A8) of \citet{stawarz08} for the boundary condition of $\tilde{N}(\gamma,1)\equiv 0$ gives the solution,
\begin{equation}
\tilde{N}(\gamma,\xi)={1\over \gamma^{2} r_{\rm j}(\xi)^{2/3}}\!\int_{\xi_{\rm m}(\gamma,\xi)}^\xi \!\!\!\!\!\!\!\!\!\!{\rm d}\xi' \tilde{Q}[\gamma'(\gamma,\xi,\xi'),\xi'] \gamma'(\gamma,\xi,\xi')^{2} r_{\rm j}(\xi')^{2/3}\!\!.
\label{A8}
\end{equation}
From equation (\ref{gamma_xi0}), $\xi_{\rm m}$ is the maximum of 1 and the solution of
\begin{equation}
\gamma\int_{\xi_{\rm m}(\gamma,\xi)}^{\xi} {\rm d}\xi' c_1(\xi')  U(\xi') [r_{\rm j}(\xi')/r_{\rm j}(\xi)]^{-2/3}=1.
\label{xm0}
\end{equation}
If $\xi_{\rm m}>1$, the distribution at $\xi$ does not depend on the conditions at $\xi<\xi_{\rm m}$. Also, $\tilde{N}(\gamma,\xi)$ depends only on the injection and cooling rate at $\gamma'>\gamma$, $\xi'<\xi$. 

Then, we consider the specific case of a conical jet with a constant speed, optically-thin synchrotron at $\gamma\gg 1$, no Compton losses, and $B\propto\xi^{-1}$, for which
\begin{eqnarray}
\lefteqn{
\gamma_\xi(\gamma,\xi)=- {2\gamma\over 3\xi}- {2 A_{\rm S}\over 3 A_{\rm ad}} {\gamma^2\over \xi^2},
\label{gamma_xi1}}\\
\lefteqn{
\gamma'(\gamma,\xi;\xi')={\gamma v^{-2/3}\over 1+{2\gamma\over 5\gamma_{\rm b}(\xi)} \left(1-v^{-5/3}\right)},\quad \gamma_{\rm b}\equiv {A_{\rm ad}\xi\over {A}_{\rm S}},\quad v\equiv {\xi'\over \xi},
\label{gamma_xi}}\\
\lefteqn{
\xi_{\rm m}(\gamma,\xi)=\cases{1,&$\gamma\leq \gamma_{\rm adv}$;\cr
\xi \left(1+{5\gamma_{\rm b}\over 2\gamma}\right)^{-3/5}\!\!\!,&$\gamma>\gamma_{\rm adv}$,}\quad \gamma_{\rm adv}\equiv {5A_{\rm ad}\xi\over 2 A_{\rm S}(\xi^{5/3}-1)},
\label{xi_m}}
\end{eqnarray}
where $\gamma_{\rm b}$ is the break Lorentz factor at which the steepening of the electron distribution due to radiative cooling occurs [though, in the approximation to the loss rate of equation (\ref{gamma_xi1}), it can be $<1$], and $\gamma_{\rm adv}$ is the Lorentz factor at which a break occurs due to advection. This break occurs due to the presence of the boundary at $\xi=1$, which changes the formula for the lower integration limit in equation (\ref{A8}) at $\gamma_{\rm adv}$. 

As stated above, inclusion of radiative cooling causes values of $\gamma$ at some $\xi$ to become independent of those at low values of $\xi'$. This is illustrated for the present case in Fig.\ \ref{gamma_vs_xi}. We see that a large enough value of $\gamma(\xi)$, i.e., above the converging line in Fig.\ \ref{gamma_vs_xi}, cannot correspond to any $\gamma'$ at $\xi'=1$, and instead has to be due to the evolution of electrons injected at $\xi' >\xi_{\rm m}>1$. 

\begin{figure*}
\centerline{\includegraphics[height=0.8\columnwidth]{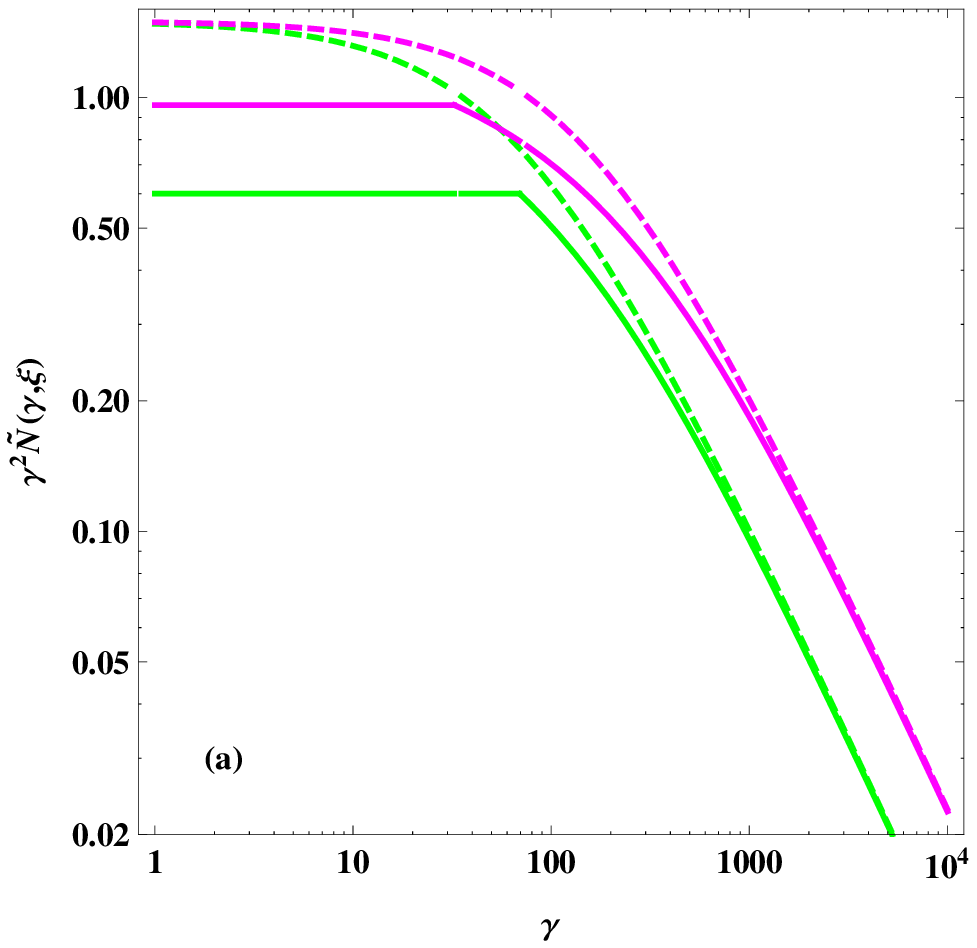} \includegraphics[height=0.8\columnwidth]{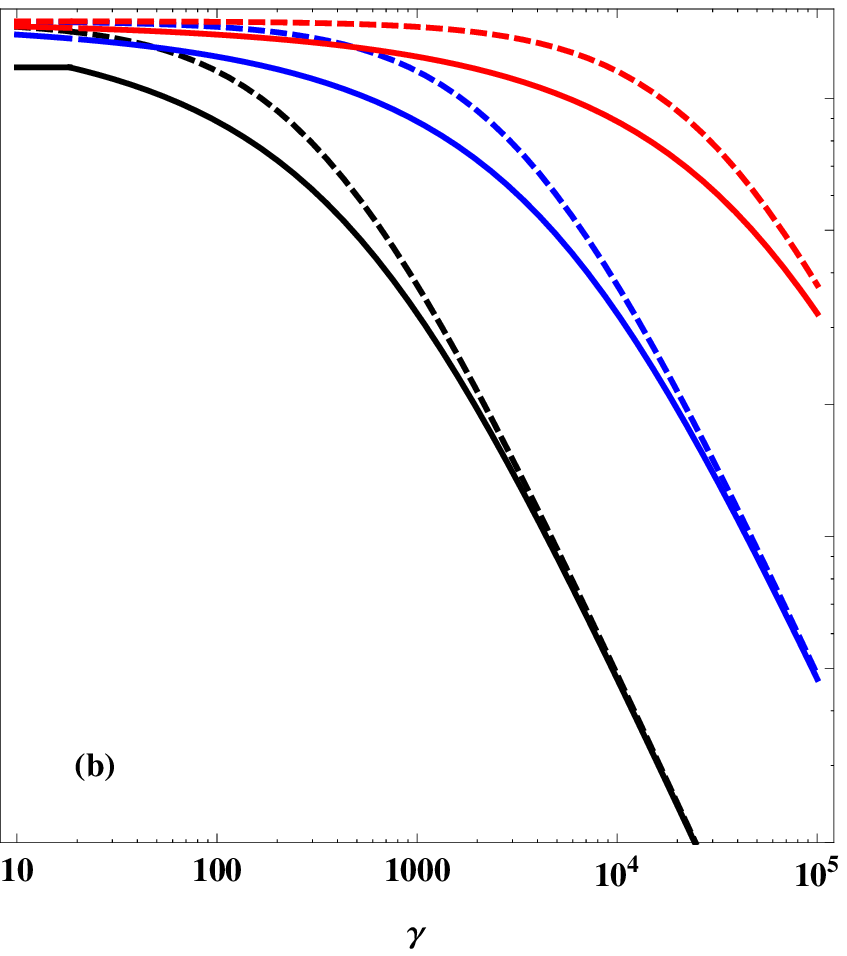}} 
\centerline{\includegraphics[height=0.8\columnwidth]{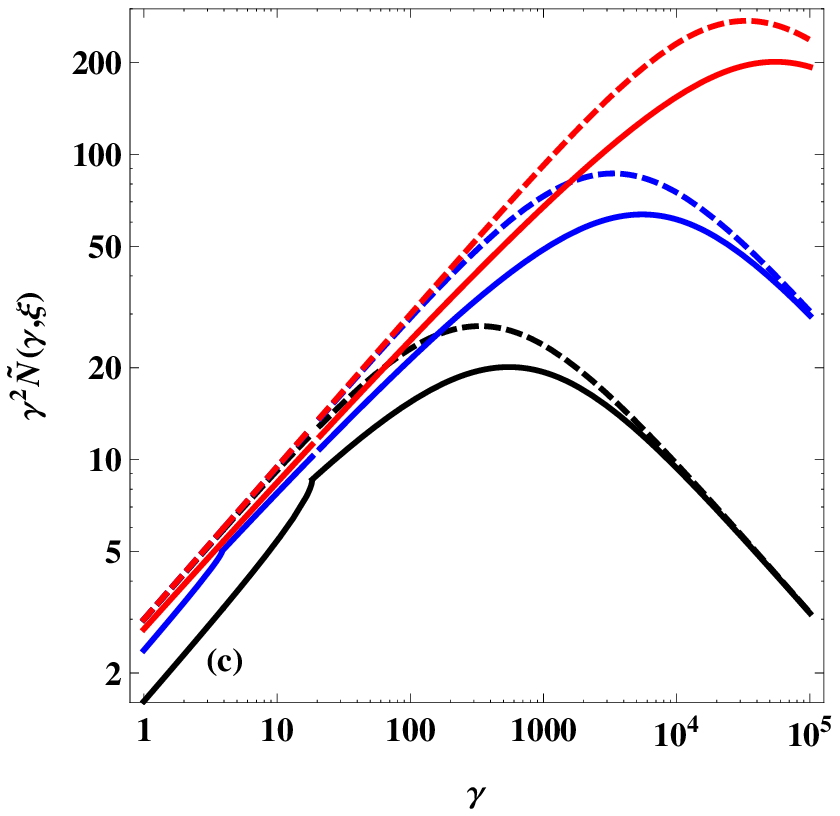} \includegraphics[height=0.8\columnwidth]{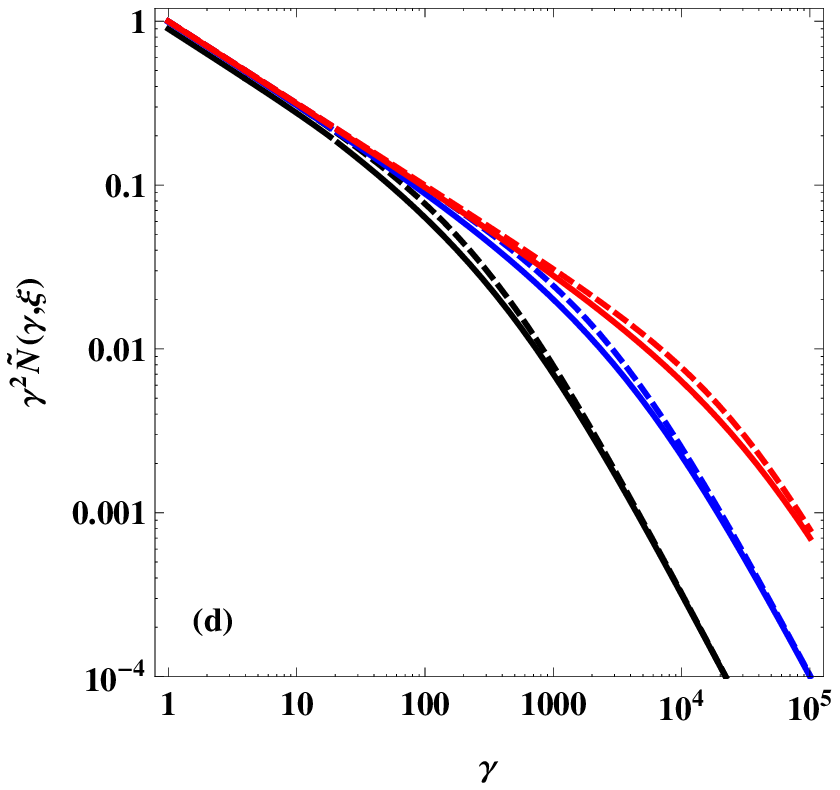}} 
\caption{Comparison of the exact solutions to the advection-losses kinetic equation (solid curves) with the local solutions (dashed curves), for $A_{\rm S}/ A_{\rm ad}=0.03$. The index of the injected electrons is (a--b) $p=2$, (c) $p=1.5$, (d) $p=2.5$. The green and magenta (shown only for $p=2$) and black, blue, red curves correspond to $\xi=10^{1/3}$, $10^{2/3}$, $10^1$, $10^2$, $10^3$, respectively. The values of $\gamma_{\rm b}$ are $\simeq 33\xi$, and those of $\gamma_{\rm adv}$ are at the positions of the kinks in the distributions. The normalization corresponds to $\tilde{Q}_0=1$.
} \label{el_dist}
\end{figure*}

We also consider in more detail a power-law injection, equation (\ref{Q_inj1}), for $q(\xi)=Q_0 \xi^{-3}$, 
\begin{eqnarray}
\lefteqn{
\tilde{Q}(\gamma,\xi)= \cases{0,&$\gamma<\gamma_{\rm m}$;\cr
\tilde{Q}_0 \xi^{-1} \gamma^{-p}g_{\rm cut}(\gamma, \gamma_{\rm M}), &$\gamma\geq \gamma_{\rm m}$,\cr}\label{Q_inj}}\\
\lefteqn{ 
\tilde{Q}_0 = {\upi z_{\rm m}^4 \tan^2 \Theta_{\rm j}\over c}Q_0.\label{Q0}}
\end{eqnarray}
For $\gamma>\gamma_{\rm m}$, $p>1$, and neglecting the high-energy cutoff, equation (\ref{A8}) can be integrated to
\begin{eqnarray}
\lefteqn{
\tilde{N}(\gamma, \xi) = {3 \tilde{Q}_0 \gamma^{-p} \over 2 (p-1)}  
\times\nonumber}\\
\lefteqn{{(\xi'/\xi)^{{2 \over 3} (p-1)}  \over \left(1+{2\gamma\over 5\gamma_{\rm b}(\xi)}\right)^{2-p}} \left.
{_2}F_1 \left[2-p, {2 - 2p\over 5}, {7-2 p \over 5}; {(\xi'/\xi)^{-5/3}  \over 1 + {5\gamma_{\rm b}(\xi)\over 2\gamma}}\right]\right|^{\xi'=\xi}_{\xi'=\xi_{\rm m}},
\label{Nint2}}
\end{eqnarray}
where ${_2}F_1$ is the Gauss hypergeometric function. When ${A}_{\rm S}=0$, i.e., there are only adiabatic losses, the solution simplifies to
\begin{equation}
\tilde{N}(\gamma,\xi)= {3\tilde{Q}_0 \gamma^{-p}\over 2(p-1)}\left(1- \xi^{-{2(p-1)\over 3}}\right).
\label{N_ad}
\end{equation}
When $p=2$, the solution with optically-thin synchrotron losses simplifies to
\begin{eqnarray}
\lefteqn{
\tilde{N}(\gamma,\xi)= {3\tilde{Q}_0 \gamma^{-2}\over 2}\left(1- v_{\rm m}^{2/ 3}\right)\nonumber}\\
\lefteqn{\quad
= {3\tilde{Q}_0 \gamma^{-2}\over 2}
\cases{\left(1- \xi^{-2/ 3}\right),&$\gamma\leq \gamma_{\rm adv}(\xi)$;\cr
\left[1- \left(1+{5\gamma_{\rm b}\over 2\gamma} \right)^{-2/5}\right], &$\gamma>\gamma_{\rm adv}(\xi)$,}\quad v_{\rm m}\equiv {\xi_{\rm m}\over \xi}.
\label{N_p2}}
\end{eqnarray}
Note that the case of $\gamma\leq \gamma_{\rm adv}$ is purely adiabatic, of equation (\ref{N_ad}). Except for $\xi\sim 1$, the case of $\gamma> \gamma_{\rm adv}$ steepens to $\propto \gamma^{-3}$ in the cooled regime,
\begin{equation}
\tilde{N}(\gamma,\xi)\simeq {3\over 2}\tilde{Q}_0 \gamma_{\rm b}\gamma^{-3}, \quad \gamma\gg\gamma_{\rm b}.
\label{N_p2_high}
\end{equation}
Another case of interest is $p=3$, for which the solution is
\begin{equation}
\tilde{N}(\gamma,\xi)={3\tilde{Q}_0 \gamma^{-3}\over 4} \left[1-v_{\rm m}^{4/3}+{2\gamma\over 5\gamma_{\rm b}}\left(5-4v_{\rm m}^{-1/3}-v_{\rm m}^{4/3}\right)\right].
\label{N_p3}
\end{equation}
The hypergeometric function in equation (\ref{Nint2}) encounter complex infinity at $p=7/2$; although the result after the subtraction is finite, this causes a numerical difficulty. To avoid this problem, equation (\ref{A8}) can be analytically integrated for this $p$, yielding a relatively simple formula.

The exact solutions of the kinetic equation discussed above can be compared to the local solution of equation (\ref{ndotc}), i.e., without the spatial advection term,
\begin{equation}
{N}_0(\gamma,\xi)={-1\over \dot{\gamma}(\gamma, \xi)} \int_\gamma^{\infty}{\rm d}\gamma'' Q(\gamma'',\xi).
\label{N_cool}
\end{equation}
The effect of the synchrotron self-absorption process can in this case be taken into account using the approximations of equation (\ref{gdot_syn0}). The equation for $N_0$ becomes then a quadratic one with the coefficients of
\begin{eqnarray}
\lefteqn{
a'=-(\dot\gamma_{\rm ad}+\dot\gamma_{\rm C})A_\tau,\quad b'=-\dot\gamma_{\rm thin}+c' A_\tau,\quad c'=-\int_\gamma^{\infty}\!\!\!\!{\rm d}\gamma'' Q,\label{abc}}\\
\lefteqn{
A_\tau={C_2 \upi \sigma_{\rm T}B_{\rm cr}z_{\rm m}\xi\tan\Theta_{\rm j} \over  2\alpha_{\rm f} B(\xi)\gamma^4},\label{A_tau}}
\end{eqnarray}
where $\dot\gamma_{\rm C}$ and $\dot\gamma_{\rm thin}$ are the energy loss rates for all the Compton processes and all the optically-thin radiative and adiabatic processes, respectively, and $C_2$ is defined below equation (\ref{alphas}). Note that while $a'>0$ and $c'<0$ always, the sign of $b'$ changes between the optically thick and thin regimes. The effect of self-absorption on $N_0$ is important if $\gamma_{\rm b}$ is $\la\gamma_{\rm t}$, the Lorentz factor corresponding to the turnover energy, equation (\ref{gt}). 

For the injection parameterized as in equation (\ref{Q_inj}) and the loss term with no synchrotron self-absorption as given by equation (\ref{gamma_xi1}), the solution is
\begin{equation}
{N}_0(\gamma,\xi)= {{Q}_0 \max(\gamma,\gamma_{\rm m})^{1-p}f_{\rm cut}[\gamma, \gamma_{\rm M}(\xi)]\over A_{\rm ad} (p-1)\xi^2 \gamma[1+\gamma/\gamma_{\rm b}(\xi)]},
\label{N_cool0}
\end{equation}
where $f_{\rm cut}$ denotes the high-energy cutoff resulting from integration of $\gamma^{-p} g_{\rm cut}$. In the adiabatic case, i.e., $\gamma_{\rm b}\gg \gamma_{\rm M}$, and for $\gamma>\gamma_{\rm m}$, this local solution is the same as the adiabatic one with advection, equation (\ref{N_ad}), except for the boundary term, $[1- \xi^{-2(p-1)/ 3}]$. Thus, except for the vicinity of $z_{\rm m}$, advection is not important for adiabatic cooling at $q\propto\xi^{-3}$. When we increase the cooling by including radiative losses, we can also expect the two solutions to be even closer to each other at high $\xi$. We confirm it numerically, and find that indeed advection plays an important role only for relatively low $\xi$ and low $\gamma$. 

Fig.\ \ref{el_dist} presents a comparison of the exact solutions to the advection-losses kinetic equation with the local solutions, for various electron injection spectra. We see that the two solutions are rather close to each other at large enough $\xi$, roughly at $\ga 10$. At lower $\xi$, the null boundary condition at $\xi=1$ causes the advective solution to be significantly lower than the cooling one at low $\gamma$. Then, the change of the formula for $v_{\rm m}$ at $\gamma_{\rm adv}$ leads to the appearance of a spectral break/kink in addition to the cooling break. The reason for it is the presence of the lower boundary of the dissipation region at $\xi=1$. We see that $\gamma_{\rm adv}$ decreases with increasing $\xi$. On the other hand, the cooling break scales with the jet height in the opposite way, namely $\gamma_{\rm b}\propto \xi$. We find $\gamma_{\rm adv}=\gamma_{\rm b}$ at $\xi\simeq 1.22$. Thus, $\gamma_{\rm adv}<\gamma_{\rm b}$ except for $\xi\sim 1$. At $\xi\geq 10$ and $p= 2$, the local solution is higher than the advective one by at most $\sim 25$ per cent, and the agreement improves with increasing $p$. Even at low values of $\xi$, the advective solution approaches the cooling one at high values of $\gamma$ (which are important for high-energy emission), due to the increasing strength of cooling causing the solution to be increasingly more local. These considerations show that for our chosen form of $Q(\gamma,\xi)$, advection represents a relatively small correction. 

In some cases, e.g., in the jet of Cyg X-1, $\dot \gamma_{\rm SSC}$ and $\dot\gamma_{\rm XC}$ are negligible \citep{z14}, but $\dot\gamma_{\rm BBC}$ is not. If $f_{\rm KN}\simeq 1$ and self-absorption is negligible, the local solution is represented by equation (\ref{N_cool0}) but with $\gamma_{\rm b}$ given by,
\begin{equation}
\gamma_{\rm b}(\xi)\simeq A_{\rm ad}\xi \left[A_{\rm S}+{A_{\rm BBC}(\xi)\over \xi^{-2}+z_{\rm m}^2/a^2}\right]^{-1}.
\label{gammab}
\end{equation}
If $1\ll A_{\rm S}/A_{\rm BBC}\ll (a/z_{\rm m})^2$ (as, e.g., in the case of Cyg X-1, \citealt{z14}), we can roughly approximate $\gamma_{\rm b}(\xi)$ as 
\begin{equation}
\gamma_{\rm b}(\xi) \simeq \gamma_{\rm b0}\cases{\xi, &$\xi\la \xi_1\equiv [A_{\rm S}/A_{\rm BBC}(\xi=0)]^{1/2}$;\cr
\xi_1^2\xi^{-1}, &$\xi_1\la \xi\la \xi_2\equiv (1-\beta_{\rm j})a/z_{\rm m}$;\cr
(\xi_1/ \xi_2)^2\xi, &$\xi_2\la \xi\leq \xi_{\rm M}\equiv z_{\rm M}/z_{\rm m}$,\cr}
\label{gb2}
\end{equation}
where $\gamma_{\rm b0}\equiv A_{\rm ad}/A_{\rm S}$, and the dominant losses in the three distinct jet regions discerned in the above are synchrotron, approximately constant BBC, and spatially diluted BBC, respectively. We see that $\tilde{N}_0$ has an approximate form of a broken power law, with $\tilde{N}_0(\gamma,\xi)\propto \gamma^{-p}$ at $\gamma\la \gamma_{\rm b}$, and $\tilde{N}_0 (\gamma,\xi) \simprop \xi\gamma^{-p-1}$ at $\gamma\ga \gamma_{\rm b}$. Note that if $\gamma_{\rm b}$ is either $<\!1$ or $>\!\gamma_{\rm M}$, the distribution becomes an approximate single power law with a cutoff. 

\subsubsection{Equipartition and jet power}
\label{eq_power}

We denote the pressure ratio between relativistic electrons and jet magnetic field by $\beta_{\rm eq}$. Then,
\begin{equation}
\beta_{\rm eq}(\xi)\equiv {u_{\rm e}(\xi)/3\over p_B}
={m_{\rm e}c^2\over 3 p_B}\int {\rm d}\gamma\,(\gamma-1) N(\gamma,\xi),\label{beta}
\end{equation}
where $u_{\rm e}$ is the energy density of the relativistic electrons, and the pressure of the magnetic field is given by
\begin{equation}
p_B=\cases{\displaystyle{B^2\over 8\upi}, & completely ordered $B$;\cr
\displaystyle{B^2\over 24\upi}, & completely tangled $B$}
\label{pb}
\end{equation}
\citep{leahy91}. 
We also calculate the magnetization parameter,
\begin{equation}
\sigma_{\rm eq}\simeq {u_B+p_B\over w}, \quad  w=u_{\rm i}+{4\over 3} u_{\rm e}\geq  n_{\rm e,rel} m_{\rm p}c^2+{4\over 3} u_{\rm e},
\label{magn}
\end{equation}
where $w$ is the proper particle enthalpy, $u_B=B^2/8\upi$ is the magnetic energy density, $u_{\rm i}$ is the ion energy density including their rest mass, $m_{\rm p}$ is the proton mass, and $n_{\rm e,rel}\equiv \int {\rm d}\gamma\, N(\gamma,\xi)$ is the number density of the accelerated electrons. Hereafter we assume cold protons with negligible pressure, and ignore the electron inertia and possible presence of e$^\pm$ pairs. The equality above corresponds to the minimum possible ion number density, equal to that of the relativistic electrons. Hence, the number and energy densities of relativistic electrons determined from observations provide the upper limit on $\sigma_{\rm eq}$. For $B\propto\xi^{-1}$ in conical jets, the minimum of $\sigma_{\rm eq}$ corresponds to the maximum of $\xi^2 n_{\rm e,rel}$, which is achieved at $\xi_{\rm M}$. 

The total rate in the observer frame at which relativistic electrons are injected into the jet+counterjet of constant velocity is,
\begin{equation}
R_{\rm inj}=2\upi z_{\rm m}^3 \tan^2\Theta_{\rm j} \int_{\xi_{\rm m}}^{\xi_{\rm M}} {\rm d}\xi\, \xi^2\! \int_{\gamma_{\rm m}}^\infty {\rm d}\gamma\, Q(\gamma,\xi).\label{Rinj}
\end{equation}
The corresponding injected power is,
\begin{equation}
P_{\rm inj}=2\upi z_{\rm m}^3 \tan^2\Theta_{\rm j} \Gamma_{\rm j}m_{\rm e} c^2 \int_{1}^{\xi_{\rm M}} {\rm d}\xi\, \xi^2\! \int_{\gamma_{\rm m}}^\infty {\rm d}\gamma\, (\gamma-1) Q(\gamma,\xi).\label{Pinj}
\end{equation}
We can compare $R_{\rm inj}$ to the flux through the jet of electrons. The density of relativistic electrons per unit jet length is maximized at $z_{\rm M}$, and thus the total electron number flux is given by,
\begin{equation}
R_{\rm e}\geq 2\upi (\xi_{\rm M}z_{\rm m}\tan\Theta_{\rm j})^2 n_{\rm e,rel}(\xi_{\rm M}) \Gamma_{\rm j}\beta_{\rm j}c,
\label{eflux}
\end{equation}
where the $\geq$ sign accounts for a possible presence of electrons other than those in the accelerated distribution. If $R_{\rm inj}>R_{\rm e}$, some of the electrons need to be accelerated more than once. This is possible since electrons injected with the distribution of $Q(\gamma,\xi)$ lose energy and may be reaccelerated. The rate (\ref{eflux}) is related to a limit on the mass flow in the jet+counterjet,
\begin{equation}
\dot M_{\rm j}\geq 2\upi (\xi_{\rm M}z_{\rm m}\tan\Theta_{\rm j})^2 n_{\rm e,rel}(\xi_{\rm M}) \Gamma_{\rm j}\beta_{\rm j}m_{\rm p} c.
\label{Mdot}
\end{equation}

The total power in a given radiative or adiabatic process is,
\begin{equation}
P_k =-2\upi z_{\rm m}^3 \tan^2\Theta_{\rm j}\Gamma_{\rm j} m_{\rm e} c^2 \int_{\xi_{\rm m}}^\infty {\rm d}\xi\, \xi^2\! \int_{\gamma_0}^\infty {\rm d}\gamma\, N(\gamma,\xi) \dot\gamma_k(\gamma,\xi),
\label{Prad}
\end{equation}
where $\dot\gamma_k$ gives either adiabatic, synchrotron, SSC, BBC or XC loss rate and $\gamma_0> 1$ is the minimum Lorentz factor to which the electrons are cooled in the emission region, which can be $<\gamma_{\rm m}$. For $\dot\gamma_k =\dot\gamma$, $P_{\rm k}= P_{\rm inj}$.

The jet+counterjet power stored in the relativistic electrons, $P_{\rm e}$, the protons (not including their rest mass), $P_{\rm i}$, and the magnetic field, $P_B$, are 
\begin{eqnarray}
\lefteqn{P_{\rm e}(\xi)=
{8\upi \over 3} u_{\rm e}(\xi) \beta_{\rm j} c (\Gamma_{\rm j} \xi z_{\rm m} \tan\Theta_{\rm j})^2,\label{pe}}\\
\lefteqn{P_{\rm i}=\dot M_{\rm j}c^2(\Gamma_{\rm j}-1)\geq\nonumber}\\
\lefteqn{\quad 
\geq 
2\upi n_{\rm e,rel}(\xi_{\rm M}) m_{\rm p}c^3 \beta_{\rm j}\Gamma_{\rm j} (\Gamma_{\rm j}-1)(\xi_{\rm M} z_{\rm m}\tan\Theta_{\rm j})^2,\label{pp}}\\
\lefteqn{P_B(\xi)=2\upi (u_B+p_B)
\beta_{\rm j} c (\Gamma_{\rm j}\xi z_{\rm m}\tan\Theta_{\rm j})^2,
\label{Pb}}
\end{eqnarray}
where $P_B$ is independent of $\xi$ for $B=B_0/\xi$. The equations above can be generalized to abundances other than pure H. E.g., for He systems (like Cyg X-3), $m_{\rm p}$ above should be replaced by $2 m_{\rm p}$.

Note that the jet kinetic power in particles can be written as
\begin{equation}
P_{\rm ei}\equiv P_{\rm e}+P_{\rm i}= A_{\rm j} Q_0\tan^2\Theta_{\rm j},
\label{power}
\end{equation}
where $A_{\rm j}$ follows from equations (\ref{pe}--\ref{pp}). This implies $Q_0\propto \tan^{-2}\Theta_{\rm j}$. Then the optically-thin part of $P_{\rm S}$, see equations  (\ref{Prad}) and (\ref{gdot_syn}), is $\propto P_{\rm ei}$, and thus independent of $\Theta_{\rm j}$. The same holds for the BBC process, which $P_{\rm BBC}\propto Q_0\tan^2\Theta_{\rm j}\propto P_{\rm ei}$. On the other hand, equations (\ref{jet_F}), (\ref{sdensity}--\ref{synsc}) below imply that $P_{\rm SSC}\propto Q_0^2\tan^3\Theta_{\rm j} \propto P_{\rm j}^2/\tan\Theta_{\rm j}$, i.e., $P_{\rm SSC}$ increases with decreasing $\Theta_{\rm j}$. 

\section{Spectra}
\label{rad_spectra}

\subsection{Synchrotron}
\label{synchrotron}

In the $\delta$-function approximation, $\epsilon \simeq \gamma^2 B/B_{\rm cr}$, the synchrotron emissivity for a broad distribution of electrons is,
\begin{equation}
j_{\rm S}(\epsilon,\xi)\simeq
{C_1(s) \sigma_{\rm T} c B_{\rm cr}^2  \epsilon^{1/2} \over 48\upi^2 }\left(B\over B_{\rm cr}\right)^{1/2} N\left(\sqrt{\epsilon B_{\rm cr}\over B},\xi \right),
\label{j_syn}
\end{equation}
where $C_1(s)$ is a (weak) function of the local power-law index, $s$, matching the corresponding ultra-relativistic formula for power-law electrons (ZLS12). Since $N$ is given by the solution of the kinetic equation including radiative cooling, we have $p\la s\la p+1$.  

The synchrotron emission is accompanied by self-absorption, which coefficient in the $\delta$-function approximation can be written as,
\begin{equation}
\alpha_{\rm S}(\epsilon,\xi)\simeq {C_2(s) \upi \sigma_{\rm T} \over  2\alpha_{\rm f}\epsilon^2} {B\over B_{\rm cr}} N\left(\sqrt{\epsilon B_{\rm cr}\over B},\xi\right),
\label{alphas}\\
\end{equation}
where $C_2(s)$ is a (weak) function of $s$ matching the corresponding ultra-relativistic formula for power-law electrons, analogously to the case of emission. 

Hereafter, we assume a conical jet with constant speed and conserved magnetic energy flux, equation (\ref{conical}). We define the optical depth along the line of sight through the jet or counterjet,
\begin{equation}
\tau_{\rm S}(\epsilon,\xi,\psi)= {2\alpha_{\rm S}(\epsilon) z_{\rm m}\xi\tan\Theta_{\rm j} \over {\cal D}_{\rm j,cj}\sin i}(1-\psi^2)^{1/2},
\label{tausyn}
\end{equation}
where $\psi\equiv x/(z\tan\Theta_{\rm j})$, $x$ is a coordinate perpendicular to $z$, and we note that $\sin(\upi-i) =\sin i$ for the counterjet. As a simplification, we calculate here $\tau_{\rm S}$ along its path using the values of $N$ and $B$ at the height $\xi$, which is a good approximation for narrow jets. 

In the case of the local solution, equation (\ref{N_cool}), the maximum of $\tau_{\rm S}$ is reached at $\xi=1$. However, the solution for $N$ including advection is obtained with the null boundary condition at $\xi=1$, and thus $\tau_{\rm S}(\xi=1)$ is also null. The maximum is then achieved at slightly higher values of $\xi$. For example, for the advective solution with dominant adiabatic losses, equation (\ref{N_ad}), $\tau_{\rm S}$ is maximized at a $\xi_\tau=[(7 p+8)/ (3 p+12)]^{3/[ 2(p-1)]}\simeq 1.25$--1.41, for $p=3.5$--1.5. Addition of radiative losses reduces $\xi_\tau$. Thus, we find numerically $\xi_\tau$ for solutions with advection, while $\xi_\tau=1$ for local solutions.

Note that due to the relativistic length transformation, $\tau_{\rm S}$ is different for the jet and counterjet even for the same $\epsilon$. We thus define the turnover energy, $\epsilon_{\rm t0}$, for the jet\footnote{Notice that the treatment of the counterjet emission in ZLS12 contains some inaccuracies due to their neglect of the difference of the observed turnover energy between the jet and counterjet.},
\begin{equation}
E_{\rm t0}={\cal D}_{\rm j}\epsilon_{\rm t0} m_{\rm e}c^2,\quad 
\tau_{\rm S}(\epsilon_{\rm t0},\xi_\tau,0)=1,
\label{eturn}
\end{equation}
based on an observed value of $E_{\rm t0}$.

The jet synchrotron flux taking into account both emission and absorption can be expressed as an integral over the synchrotron source function, $j_{\rm S}/\alpha_{\rm S}$, see \citet{heinz06} and ZLS12. However, we need now to take into account that the electron distribution, $N$, is not a single power law. As follows from equations (\ref{j_syn}--\ref{alphas}), $j_{\rm S}/\alpha_{\rm S}$ is independent of $N$. Thus, we only need to modify $\alpha_{\rm S}$ to take into account the actual form of $N$. This results in the synchrotron flux from a conical jet or counterjet,
\begin{eqnarray}
\lefteqn{
F_{\rm S}(E)=\left(m_{\rm e} c\over h\right)^3 {c C_1 z_{\rm m}^2 \tan\Theta_{\rm j} {\cal D}_{\rm j,cj}^3 \epsilon^{5/2}\sin i\over 3 C_2  D^2}\left(B_{\rm cr}\over B_0\right)^{1/2} \times\nonumber}\\
\lefteqn{
\int_{\max\left(1,{B_0\gamma_0^2\over B_{\rm cr}\epsilon}\right)}^{\xi_{\rm M}} {\rm d}\xi\,\xi^{3/2}\!\!  \int_{-1}^1 {\rm d}\psi \left\{1-\exp\left[-\tau_{\rm S}(\epsilon,\xi,\psi)\right]\right\},}
\label{thin_thick}
\end{eqnarray}
where $\tau_{\rm S}$ is given by equation (\ref{tausyn}). If $N$ is a single power law, $\tau_{\rm S}$ is given by the simple form used by ZLS12. The photon energies $E$ and $\epsilon$ are related by equation (\ref{rel}). (Note that the variable $\xi$ in ZLS12 $=\xi\epsilon/\epsilon_{\rm t0}$ in our notation.) The second term of the lower limit of the integral (\ref{thin_thick}) expresses the condition of $\gamma>\gamma_0$ of the emitting electrons. It implies that photons at some low energies can be emitted only high up in the jet (in the $\delta$-function approximation to the synchrotron emissivity). Hereafter, the transformation from the jet frame to the observer's frame is for a continuous jet, see \citet{s97}. 

Based on equation (\ref{thin_thick}), we can also determine the observed brightness temperature. We define it through the relation ${\rm d}F(E)=I(E){\rm d}\Omega$, where now ${\rm d}\Omega={\rm d}z\,{\rm d}x \sin i /D^2$ and $I$ is the specific intensity, which we set equal to the blackbody intensity at the temperature $T_{\rm b}$ in the Rayleigh-Jeans limit. This yields the brightness temperature in units of $m_{\rm e}c^2$ of
\begin{equation}
\Theta_{\rm b}(E,\xi,\psi)={C_1 \over C_2}{ {\cal D}_{\rm j,cj}\gamma(\epsilon)\over 6}
\left\{1-\exp\left[-\tau_{\rm S}(\epsilon,\xi,\psi)\right]\right\},
\label{brightness}
\end{equation}
where $\gamma=(\epsilon B_{\rm cr}\xi/B_0)^{1/2}$ is the Lorentz factor (for $\gamma\gg 1$) corresponding to synchrotron emission at $\epsilon$. We can see that the maximum $\Theta_{\rm b}$ is reached close to the local turnover energy, $\epsilon_{\rm t}(\xi)$, defined by $\tau_{\rm S}(\epsilon_{\rm t},\xi,0)=1$. The Lorentz factor corresponding to $\epsilon_{\rm t}$ is given by
\begin{equation}
\gamma_{\rm t}(\xi)=\left[B_{\rm cr}\epsilon_{\rm t}(\xi) \over B(\xi) \right]^{1\over 2},
\label{gt}
\end{equation}
and $\gamma_{\rm t0}\equiv \gamma_{\rm t}(\xi_\tau)$. Since the synchrotron energy loss is suppressed below $\gamma_{\rm t}$, equation (\ref{gdot_syn}), $N(\gamma,\xi)$ will be adiabatically cooled in most cases, $N(\gamma,\xi) \propto \xi^{-2}\gamma^{-p}$, which, with $B\propto \xi^{-1}$, imply $\epsilon_{\rm t} \propto \xi^{-1}$ and $\gamma_{\rm t}(\xi)=\gamma_{\rm t0}={\rm constant}$. With these relations, we find that at $\psi=0$, the maximum of $\Theta_{\rm b}$ is achieved at $\gamma\simeq (0.83$--$0.86)\gamma_{\rm t}$ for $p=1.3$--4.0, for which $(C_1/C_2)(\gamma/6) \left[1-\exp\left(-\tau_{\rm S}\right)\right]\simeq (0.39$--$0.095)\gamma_{\rm t}$. 

When $E/E_{\rm t0}\gg 1$, the effect of absorption can be neglected. An optically thin steady-state flux from any process for a continuous conical jet or counterjet with constant speed is given by,
\begin{equation}
F(E)={\upi {\cal D}_{\rm j,cj}^2 \tan^2\Theta_{\rm j} z_{\rm m}^3\over m_{\rm e} c^2 D^2} \int_1^\infty {\rm d}\xi\, \xi^2 j(\epsilon,\xi),
\label{jet_F}
\end{equation}
where $j$ can also depend on the direction of emission. The above equation can be obtained, e.g., by expanding equation (18) of ZLS12 for $\tau\rightarrow 0$. 

Equations (\ref{thin_thick}), (\ref{jet_F}) readily yield vertical profiles of the synchrotron flux at a given energy. In the optically thin case, we use $j(\epsilon,\xi)$ and $N(\gamma,\xi)$ of equations (\ref{j_syn}) and (\ref{N_cool0}), respectively. The resulting slopes in three regimes are (for our assumed $B\propto\xi^{-1}$)
\begin{equation}
{{\rm d}F_{\rm S}(E)\over {\rm d}\ln\xi} \propto \cases{\xi^{5/2},&self-absorbed;\cr
\xi^{(1-p)/2}, &optically-thin uncooled;\cr
\xi^{1-p/2},&optically-thin cooled.}
\label{vertical_F}
\end{equation}
 
The electrons at $\gamma\la \gamma_{\rm t}$ might approach a quasi-Maxwellian distribution due to equilibrium of synchrotron emission and absorption \citep*{dbs89,ghs98,katarzynski06,bmm08,vp09}. However, that distribution is not achieved in the jet case since the electrons lose substantial energy adiabatically, and, depending on the parameters, in Compton upscattering. Thus, we do not consider that effect. 

\subsection{Compton}
\label{compton}

Optically-thin steady-state flux from Compton scattering in a continuous conical jet or counterjet is given by equation (\ref{jet_F}). To calculate the emissivity due to this process, we need the electron distribution and the density of seed photons. 

In the case of the SSC process, we need to determine the density of the synchrotron photons within the jet. Even if the magnetic field is uniform and tangled at a given $z$, implying the synchrotron emission is locally uniform and isotropic, the synchrotron density will depend on the radius. However, we make a simplifying assumption that the density is uniform at a given $z$, and given by
\begin{equation}
n_{\rm S}(\epsilon_0,\xi)\simeq {4\upi \tan\Theta_{\rm j} \xi z_{\rm m} j_{\rm S}(\epsilon_0,\xi)\over \epsilon_0 m_{\rm e} c^3 [1+\tau_\perp(\epsilon_0,\xi)]}.
\label{sdensity}
\end{equation}
where $\tau_\perp$ is given by equation (\ref{tau_syn}). Then, the SSC emission can be calculated exactly, using the KN formula for isotropic seed photons of \citet{jones68}. We use the notation of equation (17) of \citet{zp13}, which yields for the first-order emission,
\begin{eqnarray}
\lefteqn{
j_{\rm SSC}(\epsilon,\xi)={3\sigma_{\rm T}m_{\rm e}c^3\over 16\upi}
\int_{\gamma_0}^\infty {\rm d}\gamma {N(\gamma,\xi)\over \gamma^2}
\int_{\epsilon_{\rm n}}^\infty {\rm d}\epsilon_0 {n_{\rm S}(\epsilon_0,\xi) \over \epsilon_0}
 \bigg[ 1+ 2\epsilon\epsilon_{\rm n} +
\nonumber}\\
\lefteqn{
\quad  +{\epsilon_{\rm n}(1-2\epsilon\epsilon_{\rm n})\over \epsilon_0} -{2\epsilon_{\rm n}^2\over \epsilon_0^2}+{2\epsilon_{\rm n}\over \epsilon_0}\ln{\epsilon_{\rm n}\over \epsilon_0}
\bigg],\quad \epsilon_{\rm n}\equiv {\epsilon\over 4\gamma (\gamma-\epsilon)}. \label{synsc}}
\end{eqnarray}
To get all orders of Compton emission, we can replace $n_{\rm S}$ by $n$ including the SSC photons, and solve this equation iteratively. Alternatively, we can calculate the SSC emissivity less accurately in a $\delta$-function approximation with the Thomson limit imposed as a step function. For that, we can use, e.g., equation (8) of ZLS12. 

For the BBC and XC emission, we need to take into account the angular dependence of the emission (e.g., \citealt{jackson72,dch10b}). For BBC emission and scattering of the central X-ray source, we assume the seed photons arriving from a single direction, see Section \ref{loss_rates}. The KN rate for monoenergetic seed photons scattered from one direction into another is given by \citet{aa81}. We then integrate over the seed photon and electron distributions,
\begin{eqnarray}
\lefteqn{
j(\epsilon,\xi)={3\sigma_{\rm T}m_{\rm e}c^3\over 16\upi}
\int_{\gamma_0}^\infty {\rm d}\gamma {N(\gamma,\xi)\over \gamma^2}
\int_{\epsilon_{\rm m}}^\infty {\rm d}\epsilon_0 {n_0(\epsilon_0,\xi) \over \epsilon_0}\bigg[1+\epsilon \epsilon_{\rm m} \chi+  \nonumber}\\
\lefteqn{\quad
-{2\epsilon_{\rm m}\over \epsilon_0}+{2\epsilon_{\rm m}^2\over \epsilon_0^2}\bigg],
\quad \epsilon_{\rm m}\equiv {\epsilon\over 2 \chi\gamma (\gamma-\epsilon)},\quad \chi\equiv 1-\cos\vartheta, \label{compton_aniso}}
\end{eqnarray}
where $n_0$ denotes the distribution of either blackbody or X-ray photons, given by equations (\ref{bb}) and (\ref{density_X}), respectively, and $\vartheta$ is the scattering angle in the jet frame. This scattering rate at a given $\gamma$ can be analytically integrated over the blackbody spectrum, as given by equation (15) of \citet{zp13}. In the case of scattering of disc photons, we also need to integrate ${\rm d}n_{\rm d}/{\rm d}\Omega$ over the solid angle subtended by the accretion disc, see Section \ref{loss_rates}. 

The scattering angle in the jet (or counterjet) frame is given by
\begin{equation}
\chi =\cases{ {\cal D}_{\rm j,cj} {\cal D}_* \left(1+{a\over r}\cos\phi_{\rm b}\sin i\mp {z\over r}\cos i\right),&blackbody;\cr
{\cal D}_{\rm j,cj} {\cal D}_{\rm d} \left(1+{r_{\rm d}\over r_{\rm dj}}\cos\phi\sin i\mp {r_{\rm d}\over r_{\rm dj}}\cos i\right),&disc;\cr
{\cal D}_{\rm j,cj} {\cal D}_{\rm X}(1\mp \cos i), &X-ray source,}
\label{chi}
\end{equation}
where the $-$ and $+$ sign is for the jet and counterjet, respectively, and the product of the Doppler factors was originally introduced by \citet*{dch10a}. The blackbody-integrated rate can be then integrated numerically over $N(\gamma,\xi)$. In all cases, we integrate $j$ over the jet length, equation (\ref{jet_F}). The SSC and XC emission are independent of the orbital phase. To get the time-average BBC emission, we integrate it over $\phi_{\rm b}/2\upi$. 

Appendix \ref{Thomson} gives an analytical estimate of the average BBC flux in the case of Thomson-limit emission of power-law electrons. Appendices \ref{syn_loss} and \ref{compton_loss} give some analytical estimates of the synchrotron and Compton emission in the Thomson limit taking into account the effect of cooling on the electron distribution.

\section{Photon-photon pair production}
\label{pair}

The jet spectrum will be attenuated in pair-producing photon-photon collisions, $\gamma\gamma\rightarrow {\rm e}^+ {\rm e}^-$. There are several sources of photons potentially absorbing \g-rays. One is accretion, which takes place close to the binary plane, and which emission consists mostly of disc blackbody and Comptonization in a hot plasma [see Section \ref{distribution}, equations (\ref{flux_X}--\ref{density_X})]. This has been considered, e.g., by \citet*{zmb09}, who found that the optical depth to this process in Cyg X-1 is $\la 1$ at any energy for $z\ga 10^9$ cm, see their fig.\ 3. In the case of AGN, pair production on disc photons has been considered, e.g., by \citet{dermer09}. Other photon sources are synchrotron and SSC (e.g., \citealt{gt09}). In the case of AGN, we also need to consider absorption on photons from broad-line regions, and cosmic microwave and stellar background.

An important contribution to opacity in high-mass X-ray binaries is that from stellar photons. Pair production on those photons is important, e.g., in Cyg X-1, see \citet{bg07}, \citet*{bka08}, \citet{zmb09}, \citet*{rdo10}. We treat it using the formalism described in Appendix \ref{taupair}. The optical depth, $\tau_{\gamma\gamma}$, strongly depends on both the orbital phase and the height along the jet. Furthermore, the BBC component is emitted anisotropically and with the emissivity strongly dependent on the height. We calculate the effect of pair absorption exactly, integrating the attenuated BBC emission over both the height and phase. On the other hand, the SSC and XC fluxes are independent of the orbital phase. Then, the SSC emission for $p\geq 2$ and the XC radiation are emitted mostly at $z\ll a$, and thus it can be calculated at the jet base, neglecting the dependence of $\tau_{\gamma\gamma}(z)$.

We note that the absorbed \g-rays will produce pairs, which, in turn, may Compton upscatter the stellar radiation, thus initiating a spatially-extended pair cascade, see, e.g., \citet{bednarek97}, \citet{bg07}, \citet{bka08}, \citet{zmb09}. This will give rise to emission at energies lower than those of the absorbed \g-rays. However, for magnetic fields expected around a supergiant, the pairs may predominantly emit synchrotron (rather than Compton) emission \citep{bka08}. The synchrotron emission of the pairs is then negligible compared to other emission of the system. We neglect this effect in our treatment of pair absorption.

\section{Model parameters and solutions}
\label{solutions}

We have provided a set of equations allowing us to calculate emission of an extended jet with specified shape, magnetic field strength vs.\ height, and the rate of acceleration of relativistic electrons as a function of the Lorentz factor and height. Some of our results are general, but most follow the assumptions given in equation (\ref{conical}), i.e., the jet is conical, has a constant speed, the magnetic energy flux is conserved, and the power injected per unit logarithmic height is constant. Also, following equation (\ref{Q_inj}), the electrons are accelerated with a power-law distribution above some minimum Lorentz factor and with a high-energy cutoff given by the balance of acceleration and synchrotron losses, equation (\ref{gemax}). If we accept those assumptions, we can choose the parameters based on observational (e.g., radio maps, studies of binary parameters) and theoretical arguments: $\Theta_{\rm j}$, $\beta_{\rm j}$, $z_{\rm M}$, $i$ and $D$.

The six free parameters of a model are then $\eta_{\rm acc}$, $Q_0$, $p$, $z_{\rm m}$, $B_0$, and $\gamma_{\rm m}$, which can be fitted to an observed spectrum. A value of $\gamma_{\rm m}$ can be chosen within some limits on the basis of theoretical considerations, see Section \ref{acceleration}. If we have a broad-band spectrum of a jet, we can find the best fit of the parameters, and see how well the model describes the data. On the other hand, we need a minimum of five observables to determine the remaining five free parameters. As an example, we can take them as the flux in the self-absorbed part of the synchrotron spectrum, $F_{\rm thick}(E_{\rm thick})$, the turnover energy, $E_{\rm t0}$, the energy of the high-energy cutoff in the optically-thin synchrotron spectrum, $E_{\rm M}$, an optically-thin synchrotron flux, $F_{\rm thin}(E_{\rm thin})$, and the flux at a \g-ray high energy, $F_\gamma (E_\gamma)$. The solution can then be obtained using the following main relations. 
\begin{enumerate}
\item Equation (\ref{gemax}), which relates $\eta_{\rm acc}$ to $E_{\rm M}$;
\item equation (\ref{eturn}) at $E_{\rm t0}$, requiring $\tau_{\rm S}(E_{\rm t0})=1$;
\item equation (\ref{thin_thick}) for the synchrotron flux at $E_{\rm thick}$, requiring it equals $F_{\rm thick}$;
\item equation (\ref{jet_F}) for the flux at $E_{\rm thin}$, requiring it equals $F_{\rm thin}$;
\item equations (\ref{synsc}--\ref{compton_aniso}) at $E_\gamma$, requiring the sum of the BBC, SSC and XC fluxes averaged over the orbital phase and photon-photon pair-absorbed equals $F_\gamma$.
\end{enumerate}
We can solve these equations simultaneously or by iteration. Generally, the relation (i) yields $\eta_{\rm acc}$, (ii) gives an estimate of $Q_0$, (iii) relates $B_0$ to $z_{\rm m}$, (iv) $F_{\rm thin}$ strongly depends on $p$, and (v) $F_\gamma$ is strongly anti-correlated with $B_0$. In the absence of one of the observables, we can assume some form of equipartition, either $\beta_{\rm eq}\sim 1$ or $\sigma_{\rm eq}\sim 1$. We can either use the exact formalism including electron advection, Section \ref{kinetic}, or approximate $N(\gamma,\xi)$ by its local form, equation (\ref{N_cool}). 

If we have a solution, we can calculate its implied values of the equipartition parameters, $\beta_{\rm eq}$ and $\sigma_{\rm eq}$. These parameters define three regimes: (i) $\beta_{\rm eq}\ga 1$, electron pressure dominates over that of the magnetic field, (ii) $\beta_{\rm eq}\la 1$ and $\sigma_{\rm eq}\la 1$, magnetic pressure dominates over that of the electrons but is not important for the jet dynamics, and (iii) $\sigma_{\rm eq}\ga 1$, the magnetic field dominates the dynamics. Any efficient diffusive shock acceleration is expected to take place in the weak-magnetization regime, $\sigma_{\rm eq} < 1$. 

Given the components of the jet power, Section \ref{eq_power}, we check their physical self-consistency. For example, our assumption of a constant jet speed requires that the power injected in relativistic electrons, $P_{\rm inj}$ (approximately equal to the total radiated power), is $\ll$ the jet power in particles and magnetic field, $P_{\rm e}+P_{\rm i}+P_B$. The radiated power can be at the expense of the power in ions, and can be dissipated by shock acceleration, or at the expense of the magnetic field, dissipated by reconnection. 

\section{Conclusions}
\label{conclusions}

Our main results are as follows.

We have formulated a detailed model of a jet emitting in an extended height range, taking into account continuous electron acceleration along the outflow.
Extended jets with distributed energy dissipation are likely to be present in the hard spectral state of black-hole binaries. The synchrotron spectra from various heights are partially self-absorbed, and the superposition of them yields a radio spectrum with $\alpha=0$. 

Assuming power-law scaling of the electron acceleration rate, we derive the steady-state electron distribution along the jet, taking into account acceleration, adiabatic, synchrotron and Compton energy losses and advection. In the case of Compton losses being in the Thomson limit, we find the solution with advection in a form of a single integral, and we find an analytical solution for the radiative losses dominated by optically-thin synchrotron. For our general solution, we take into account the Compton losses with the Klein-Nishina cross section and blackbody, synchrotron and accretion-flow seed photons. 

We then derive self-consistently formulae for the photon spectra emitted locally by given steady-state electron distributions. The synchrotron spectra are calculated from the equation of radiative transfer, which gives the results valid from the self-absorbed to optically thin regimes. The condition of $\tau_{\rm S}=1$ together with the condition of emitting the flux as observed at the turnover energy relates the height and the magnetic field strength at the place of the onset of the jet emission. 

Compton scattering in jets is usually in the optically-thin regime, i.e., only single scattering is important. Compton spectra are derived for blackbody, synchrotron and accretion-flow seed photons. At high energies, the locally emitted spectra are attenuated by photon-photon pair production on stellar photons, taking into account the finite extent of the star. We then integrate the spectra over the jet height. Both Compton scattering on stellar photons and pair absorption depend on the orbital phase. To derive the orbit-average flux, the spectra need to be integrated over the phase.

Compton scattering of stellar blackbody photons by jets in the hard state of black-hole binaries with a high-mass companion was found to be important by \citet{mzc13} and \citet{z14}. Those jets are extended, and all the electrons in the jet at least up to the height of the order of the binary separation can efficiently up-scatter stellar blackbody photons. This situation is different from that for synchrotron seed photons, which emission is often dominated by the jet base. Thus, the flux from Compton upscattering of stellar photons can be quite high, and is likely to dominate the SSC flux. For standard estimates of the jet magnetic field and acceleration limited by radiative losses, this process is predicted to give substantial flux above the MeV range. We also find that Klein-Nishina effects substantially modify the electron distribution. 

We also present details of the treatment of \citet{bednarek97} of the optical depth due to photon-photon pair production in the field of a star, taking into account its finite extent. This calculation allows us to express it as a triple integral, compared to a quadruple one previously calculated.

The formalism developed in this work is applied to the hard state of Cyg X-1 in \citet{z14}. We present there detailed spectra, electron distributions and cooling rates as functions of the jet height and the spectra integrated over the jet.

\section*{ACKNOWLEDGMENTS}

This research has been supported in part by the Polish NCN grants N N203 581240, 2012/04/M/ST9/00780, ({\L}.S.) DEC-2012/04/A/ST9/00083 and (M.S.) DEC-2011/01/B/ST9/04845. We thank G. Romero, R. Moderski, W. Potter, P. Bordas and W. Bednarek for valuable suggestions and discussions, and the referee for valuable suggestions.

\appendix

\section{Adiabatic losses}
\label{adiabatic}

The presence of an external medium is often claimed to be required for adiabatic energy losses of an expanding plasma to occur. The expanding plasma then performs work on heating that medium, which causes the internal energy of the plasma to decrease. However, an adiabatic term is present in the hydrodynamic energy equation for thermal plasmas regardless of any external medium. It causes the internal energy of an outflow/inflow to decrease/increase. Two examples are compressional heating of a hot accretion flow onto a black hole, e.g., \citet{yuan01}, and adiabatic losses of stellar wind, e.g., \citet{zdz12}, which take place regardless of external media. In an outflow, the energy lost adiabatically increases the flow bulk motion, or, in an inflow, the adiabatic heating is at the cost of the bulk motion. Microscopically, when isotropically moving particles migrate into a larger volume, a part of their random velocities change into directed ones along the outflow. 

In the framework of the model considered here, the jet energy flux consists predominantly of the electron internal energy, equation (\ref{pe}), and of the bulk motion dominated by ions, equation (\ref{pp}). Neglecting for simplicity radiative losses and the dynamic part of magnetic field, the sum of these two terms is conserved. When the electrons move up, increasing the volume per electron, their isotropic part of the energy decreases, equation (\ref{gdot_ad}), and they gain a velocity component along the jet. The electrons form an MHD fluid together with protons and magnetic fields (even for very weak magnetic fields the mean free path of the charged particles, being determined by a multiple of their Larmor radii, is many orders smaller than the cross-sectional size of the flow). The energy lost adiabatically goes into the kinetic energy of the entire jet, i.e., the steady-state jet velocity increases with height. This acceleration is negligibly small if $P_{\rm i}\gg P_{\rm ad}$, where the power lost adiabatically, $P_{\rm ad}$, is given by equations (\ref{gdot_ad}) and (\ref{Prad}). In this work, we assume a constant jet velocity, but only as a simplification. If the above inequality is not satisfied, the increase of the bulk velocity with height should be taken into account. Also, if the ion sound speed is comparable to the bulk velocity, the jet opening angle will increase with height.

We note that some authors, e.g., \citet{kaiser06}, \citet{pc09} and \citet{pc12} considered adiabatic losses as optional, studying jet models both with and without adiabatic losses. Also, \citet{pc12} argued that both the bulk velocity and the electron internal energy may remain constant in a conical jet of constant speed. However, as we discuss above, electrons always lose energy when the volume per electron increases, and this lost energy goes into another form, in particular into the jet bulk motion, see, e.g., \citet{lb04}. 

\section{Compton upscattering of stellar radiation in the Thomson limit}
\label{Thomson}

The BBC emissivity for power-law electrons, stellar photons approximated as coming from a point source and in the Thomson limit is given by equation (A9) of \citet{z12} (see also \citealt*{dch10b}), with the dependence on the orbital angle $\propto \chi^{(1+p)/2}$. For a perpendicular jet, its average, $\langle j(z)\rangle$, over the orbital angle can be obtained in a closed form. Using equation (\ref{chi}), we obtain
\begin{eqnarray}
\lefteqn{
\left\langle \left(\chi\over D_*{\cal D}_{\rm j,cj}\right)^{1+p\over 2}\right\rangle =\cases{\displaystyle{\kappa ^{{1+p}\over 2}{_2}F_1\left({1\over 2},{-1-p\over 2},1,{-2a\sin i\over r \kappa }\right)},&any $p$;\cr
1+{z^2\over r^2}\cos^2 i +{a^2\over 2 r^2}\sin^2 i\mp 2{z\over r}\cos i, &$p=3$,}
\nonumber}\\
\lefteqn{
\kappa \equiv 1-{a\over r}\sin i\mp {z\over r}\cos i,}
\label{orb_av}
\end{eqnarray}
where the $-$ and $+$ signs are for the jet and counterjet, respectively. The total jet emission can be found numerically from equation (\ref{jet_F}).

We can also find a simple analytic estimate of the flux from the jet in the Thomson limit and neglecting the high-energy cutoff, advection and radiative cooling, i.e., $N(\gamma,\xi)=K_0 \xi^{-2}\gamma^{-p}$. We assume that the stellar blackbody irradiates the jet perpendicularly and the distance from the jet to the stellar centre equals $a$. This holds for the regions close to the jet origin, but we assume it is valid up to $z=a/2$, and we neglect the emission from higher regions, where the blackbody flux becomes diluted. We then obtain the jet flux, equation (\ref{jet_F}), averaged over the orbital phase, as
\begin{eqnarray}
\lefteqn{
F_{\rm BBC}(E)\simeq {2^{p-11\over 2}3\upi \sigma_{\rm T} K_0 (z_{\rm m}\Theta_{\rm j} r_*)^2 (k T_*)^3 \over c^2 h^3 a D^2}   
\left[\Gamma_{\rm j}(1- \beta_{\rm j}\cos i)\right]^{-2-p}\times  \nonumber}\\
\lefteqn{ \quad  \left(1-\sin i\right)^{{1+p}\over 2}{_2}F_1\left({1\over 2},{-1-p\over 2},1,{-2\sin i\over 1-\sin i}\right)\times  \nonumber}\\
\lefteqn{ \quad
{11+4p+p^2\over 5+p}  \Gamma\left(1+p\over 2\right)\zeta\left(5+p\over 2\right)\left(E\over k T\right)^{{1-p}\over 2},\label{base_sp}}
\end{eqnarray}
where $\Gamma$ and $\zeta$ are the Gamma and Riemann functions, respectively. In the range of $p=(2$--3.5) and $i= (20\degr$--$45\degr)$, equation (\ref{base_sp}) is accurate to within $\la 30$ per cent compared to the jet emission integrated over the height in the Thomson limit. Note that the accuracy of this approximation is independent of $\beta_{\rm j}$. We also note that $N(\gamma,\xi)\propto \xi^{-1}$ in the fully cooled regime, and thus the above formula does not apply to that case. 

Equation (\ref{base_sp}) strongly underestimates the counterjet emission, i.e., with the substitution of $-\beta_{\rm j}$ for $\beta_{\rm j}$, due to the much larger values of $\chi$ for the counterjet, see equation (\ref{chi}). However, the actual counterjet contribution is still small if the jet velocity is even moderately relativistic; e.g., for the $\beta_{\rm j}=0.6$ and $i=30\degr$, the exact Thomson jet/counterjet ratio is 20 and 43 for $p=2$ and 3, respectively. Still, the approximation (\ref{base_sp}) breaks down at low values of $\beta_{\rm j}$ due to the above reason.

\section{Radiative losses dominated by the synchrotron process}
\label{syn_loss}

Here, we consider analytical estimates of spectra resulting in cases when the radiative losses are mostly synchrotron and advection can be neglected. Then, equations (\ref{gammab}--\ref{gb2}) imply,
\begin{equation}
\gamma_{\rm b}=\gamma_{\rm b0}\xi, \quad \epsilon_{\rm b}=\epsilon_{\rm b0}\xi, 
\label{gamma_b}
\end{equation}
where $\epsilon_{\rm b0}$ synchrotron energy emitted by electrons at $\gamma_{\rm b0}$. The normalization of the emission at $\epsilon>\epsilon_{\rm b}$ has an additional factor of $\xi$, increasing the importance of high-energy emission at high $z$. Using equations (\ref{conical}), (\ref{N_cool0}), (\ref{j_syn}) and (\ref{gamma_b}), and neglecting the high-energy cutoff (which, if needed, can be included as $f_{\rm cut}[(\epsilon \xi B_{\rm cr}/B_0)^{1/2}, \gamma_{\rm M}(\xi)]$), the synchrotron emissivity can be written as
\begin{equation}
j_{\rm S}(\epsilon,\xi)=j_{\rm S}(\epsilon_{\rm b0},1)
\cases{\left(\epsilon\over \epsilon_{\rm b0}\right)^{-\alpha}\xi^{-\alpha-3},
&$ \epsilon\leq  \epsilon_{\rm b0}\xi$;\cr
\left(\epsilon\over \epsilon_{\rm b0}\right)^{-\alpha-1/2}\xi^{-\alpha-5/2},
&$ \epsilon\geq  \epsilon_{\rm b0}\xi$.\cr}
\label{syn_em}
\end{equation}
We can integrate this emissivity along the jet using equation (\ref{jet_F}),
\begin{eqnarray}
\lefteqn{
F_{\rm S}(E)\propto \nonumber }\\
\lefteqn{
\cases{
\left(E\over E_{\rm b0}\right)^{-\alpha}\int\limits_1^{\xi_1} {\rm d}\xi \,\xi^{-\alpha-1}, &$ E_{\rm t0}\leq E\leq  E_{\rm b0}$;\cr
\left(E\over E_{\rm b0}\right)^{-\alpha-{1\over 2}}\!\!\!\int\limits_1^{\min(\xi_1,{E\over E_{\rm b0}})}\!\!\!\!\!\!\!\!\!{\rm d}\xi \,\xi^{-\alpha-{1\over 2}} + \left(E\over E_{\rm b0}\right)^{-\alpha}\!\!\!\!\!\!\int\limits_{\min(\xi_1,{E\over E_{\rm b0}})}^{\xi_1}\!\!\!\!\!\!\!\!\! {\rm d}\xi \,\xi^{-\alpha-1},&$E\geq  E_{\rm b0}$,\cr}
}
\label{xi_int}
\end{eqnarray}
where $E_{\rm b0}$ is related to $\epsilon_{\rm b0}$ via equation (\ref{rel}). Here $\xi_1$, giving the maximum jet height at which equations (\ref{gamma_b}--\ref{syn_em}) apply, may be equal to $\xi_{\rm M}$ in the absence of BBC cooling. The lower and higher parts of the latter integration are over the radiatively cooled and adiabatically cooled emissivities, respectively. This yields
\begin{eqnarray}
\lefteqn{
F_{\rm S}(E)\simeq {F_{\rm S}(E_{\rm b0})\over 1-\xi_1^{-\alpha}}\times\nonumber}\\
\lefteqn{
\cases{(1- \xi_1^{-\alpha})\left(E\over E_{\rm b0}\right)^{-\alpha},& $ E_{\rm t0}\leq E\leq  E_{\rm b0}$;\cr
{1\over 1-2\alpha} \left(E\over E_{\rm b0}\right)^{-2\alpha}\!+{2\alpha\over 2\alpha-1} \left(E\over E_{\rm b0}\right)^{-\alpha-{1\over 2}}\!- \xi_1^{-\alpha}\left(E\over E_{\rm b0}\right)^{-\alpha}\!,& $1\leq {E\over E_{\rm b0}}\leq \xi_1$;\cr
{2\alpha\over 1-2\alpha}\left(E\over E_{\rm b0}\right)^{-\alpha-{1\over 2}} \left(\xi_1^{-\alpha+{1\over 2}}-1\right), &${E\over E_{\rm b0}}\geq \xi_1$.\label{xi1}\cr}}
\end{eqnarray}
At photon energies satisfying $1\leq E/E_{\rm b0}\ll \xi_1$, the contribution of the upper boundary of the upper region is negligible. Then, due to $\gamma_{\rm b}$ changing with height, the spectral steepening of the synchrotron spectrum due to radiative cooling is from $\alpha$ to $2\alpha$ at $\alpha<0.5$, and the usual one from $\alpha$ to $\alpha+1/2$ occurs only for $\alpha>0.5$. This form of the cooling steepening was first found by \citet{konigl81}. 

The effect of the upper boundary is pronounced only in the highest energy range, $E/E_{\rm b0}> \xi_1$ at $\alpha<0.5$. In that case, the flux is increasing with increasing $\xi_1$ as a power law, see equation (\ref{xi1}).

\section{Radiative losses from synchrotron and Thomson processes}
\label{compton_loss}

Here, we find some analytical estimates in the cases when both synchrotron and BBC in the Thomson limit are important processes, but advection and Klein-Nishina corrections can be neglected. The maximum Lorentz factor of the accelerated electrons, $\gamma_{\rm M}$, is only weakly affected by Compton scattering of blackbody photons. This is because $\gamma_{\rm M}$ is usually large enough for Compton scattering to be in the extreme KN limit. Then, its $\dot \gamma$ is strongly reduced with respect to the case of the Thomson scattering, see Fig.\ \ref{fkn}, and $\gamma_{\rm M}$ is still given by equation (\ref{gemax}).

We first consider the synchrotron emission. In the first spatial region of equation (\ref{gb2}), $\xi\leq \xi_1$, it is described by equation (\ref{xi1}). We then find the contribution from $\xi>\xi_1$ is important only at the highest energies, $E\ga E_{\rm b0}\xi_1$, and for $\alpha<1/2$. E.g., inclusion of the middle region gives a somewhat higher coefficient at the $\xi_1^{-\alpha +1/2}(E/ E_{\rm b0})^{-\alpha-1/2}$ term, $=8\alpha/ [(3+2\alpha)(1-2\alpha)]$. Furthermore, the upper region may actually give the highest contribution to the jet flux, though only for $\alpha<0.5$ and high values of $\xi_{\rm M}$. In comparison, the synchrotron flux is almost independent of $\xi_{\rm M}$ in the case of a power-law electron spectrum of a constant shape, see, e.g., ZLS12. 

We then consider the BBC process. As in equation (\ref{gb2}), we assume the irradiation to be constant up to $\xi_2$, and geometrically diluted, $\propto (\xi_2/\xi)^2$ at higher radii. The local emissivity in the Thomson regime follows a broken power-law form with the break energy, $\epsilon_{\rm b}$, dependent on $\xi$. The cooled emission usually dominates in the high-energy \g-ray region. The local $j$ can be then written as,
\begin{eqnarray}
\lefteqn{
j_{\rm BBC}\simeq j_{\rm BBC}(\epsilon_{\rm b0},1) \xi^{-2}\times \nonumber}\\
\lefteqn{\qquad
\min\left[1,\left(\xi_2\over \xi\right)^2\right]\cases{\left(\epsilon\over \epsilon_{\rm b0}\right)^{-\alpha}, &$\epsilon\leq \epsilon_{\rm b}=\epsilon_{\rm b0}\left(\gamma_{\rm b}\over \gamma_{\rm b0}\right)^2$;\cr
\left(\epsilon\over \epsilon_{\rm b0}\right)^{-\alpha-1/2}\!\!{\gamma_{\rm b}\over \gamma_{\rm b0}}, &$\epsilon\geq \epsilon_{\rm b}$,}
\label{local_j}}
\end{eqnarray}
which is continuous at $\epsilon_{\rm b0}$, and where $\epsilon_{\rm b0}\equiv \gamma_{\rm b0}^2 \bar\epsilon_{\rm bb}$, and $\bar\epsilon_{\rm bb}\equiv 2.7k T_*/({\cal D}_* m_{\rm e}c^2)$ is the average dimensionless blackbody energy in the jet frame. We then normalize the emission to the flux at $E_{\rm b0}$. Using equation (\ref{jet_F}) and integrating at $\xi\leq \xi_2$ and $\xi\geq \xi_2$, we obtain,
\begin{equation}
F_{\rm BBC}(E_{\rm b0})\simeq {2\upi {\cal D}_{\rm j}^2\Theta_{\rm j}^2 z_{\rm m}^3\xi_2 j_0\over m_{\rm e}c^2 D^2}.
\label{feb0}
\end{equation}
Then, using equations (\ref{gb2}) and (\ref{jet_F}), we find in the Thomson limit,
\begin{eqnarray}
\lefteqn{
F_{\rm BBC}(E)\simeq F_{\rm BBC}(E_{\rm b0})\times\nonumber}\\
\lefteqn{\quad
\cases{\left(E\over E_{\rm b0}\right)^{-\alpha}, &$E\leq E_{\rm b0}$;\cr
\left(E\over E_{\rm b0}\right)^{-\alpha-1/2}
{\xi_1^2\left[{1/2}+\ln(\xi_{\rm M}/\xi_1)\right]\over 2\xi_2},&$E\geq E_{\rm b0}\xi_1^2 \max\left[1,(\xi_1\xi_{\rm M})^2/\xi_2^4\right]$,}\label{bb_gb}}
\end{eqnarray}
in the fully adiabatically cooled and fully radiatively cooled regimes, respectively. There are some intermediate dependencies in between, which can be determined by integrating piecewise over the three different regimes of equation (\ref{gb2}). We note that with increasing $\xi_{\rm M}$, the fully cooled regime takes place for increasingly high energies. However, the applicability of the Thomson limit is required for the validity of the above formula, which roughly corresponds to $E\la 10^9$ eV for $T_*=3\times 10^4$ K.

\section{The optical depth for photon-photon pair production in the field of a star}
\label{taupair}

\begin{figure}
\centerline{\includegraphics[width=0.75\columnwidth]{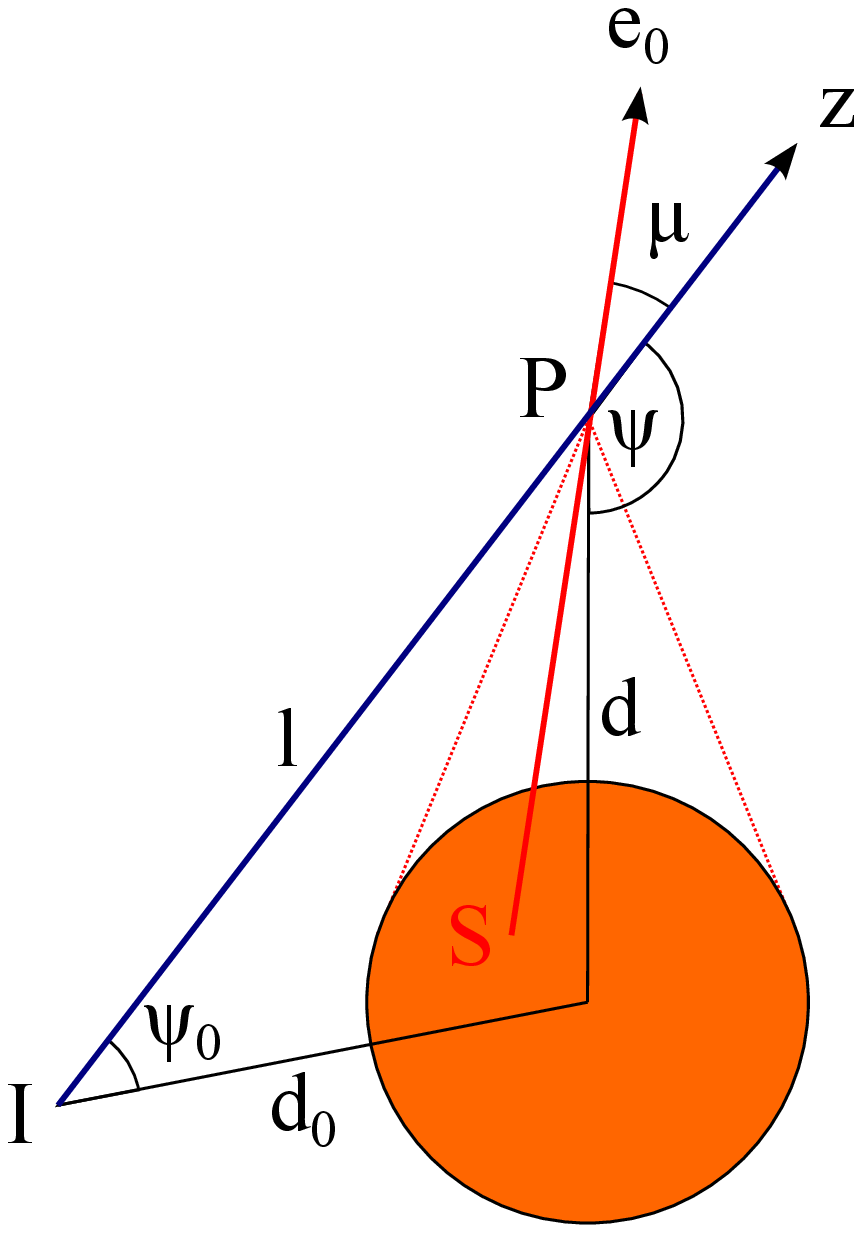}} 
\caption{Geometry of pair absorption. A \g-ray is emitted at $I$ at an angle of $\Psi_0$ with respect to the direction to the stellar centre (at the distance of $d_0$). It travels a path of $l$ before colliding with a stellar photon at $P$, at which the angle with respect to the radial direction is $\Psi$. The stellar photon is emitted at $S$ and the cosine of its direction, ${\mathbfit e_{0}}$, with respect to the \g-ray is $\mu$. The dotted lines show the range of $\mu$ corresponding to angles subtended by the star, $\mu_1$--$\mu_2$. The $z$ axis is along the \g-ray, the $x$ axis is in the plane of the drawing, and the azimuthal angle, $\phi$, is measured away from this plane.  
} \label{pair_geo}
\end{figure}

\citet{dubus06} has calculated the optical depth to absorption in pair production by a \g-ray in the field of photons from a star, $\tau_{\gamma\gamma}$, taking into account its finite extent. In his method, the optical depth is given as a four-dimensional integral, over the blackbody distribution of stellar photons, the polar and azimuthal angles of their directions, and the path of a \g-ray. As noted by \citet{bednarek97}, it is possible to integrate over the azimuth analytically with a specific choice of the coordinate system. However, \citet{bednarek97} did not provide any details, in particular the formula resulting from that integration. Then, \citet{st08} described his result. Unfortunately, their description was rather unclear and it appears difficult to implement that algorithm based on their work. Thus, we present here these results explicitly. Our geometry is shown in Fig.\ \ref{pair_geo}. The \g-ray path is in the $(x,z)$ plane. 

The differential optical depth is,
\begin{equation}
{\rm d}\tau_{\gamma\gamma}={{\rm d}n_0(\epsilon_0)\over {\rm d}\Omega} (1-\mu)\sigma_{\gamma\gamma}(\beta){\rm d}\epsilon_0 {\rm d}\mu {\rm d}\phi {\rm d}l,
\label{dtaugg}
\end{equation}
where ${\rm d}\Omega={\rm d}\mu {\rm d}\phi$ is the solid angle of the arriving soft photons, $\beta$ is velocity of the produced e$^\pm$ in the centre of momentum frame, $l$ is the path along the \g-ray, $\mu$ is the cosine of the (polar) angle between the \g-ray and blackbody photon, $\phi$ is the azimuthal angle, and $\sigma_{\gamma\gamma}$ is the pair-production cross section (\citealt{nikishov62}; see also \citealt{gs67}). The blackbody photon density per unit solid angle (i.e., the specific intensity divided by the $c\epsilon$) is given by 
\begin{equation}
{{\rm d}n_0(\epsilon_0)\over {\rm d}\Omega}=\left(m_{\rm e} c\over  h\right)^3{2\epsilon_0^2 \over \exp(\epsilon_0/\Theta_*)-1}\, {\rm cm}^{-3}{\rm sr}^{-1},
\label{bb_ster}
\end{equation}
where $\Theta_*=k T_*/m_{\rm e} c^2$. (Note that the blackbody normalization of \citealt{st08} is by a factor of $2\upi$ too large.) It is convenient to perform the integration over the blackbody photon energy changing variables from $\epsilon_0$ to $\beta$,
\begin{equation}
\epsilon_0={2\over \epsilon(1-\mu)(1-\beta^2)},\quad {\rm d\epsilon_0}={4\beta\, {\rm d}\beta\over \epsilon(1-\mu)(1-\beta^2)^2}.
\label{beta_e}
\end{equation}

A \g-ray is produced at a distance from the centre, $d_0$, at an angle, $\Psi_0$, see Fig.\ \ref{pair_geo}. These quantities are related to those at the interaction point by
\begin{equation}
d(\Psi)=d_0{\sin\Psi_0\over \sin\Psi},
\label{d_d0}
\end{equation}
where $\Psi$ is the angle between the direction of the \g-ray and the direction from the interaction point to the stellar centre and $d$ is the distance between the interaction point and the centre. The ranges of $\mu$ and $\phi$ from which blackbody photons arrive are limited by the solid angle subtended by the star as seen from the interaction point. The cosine of the polar angle is within the range $\mu_1\leq \mu\leq\mu_2$ (see the dotted lines in Fig.\ \ref{pair_geo}), 
\begin{eqnarray}
\lefteqn{
\mu_1=\cases{-1, &$\Psi_0\leq\arcsin{r_*\over d_0}$;\cr
-\left(1-{r_*^2\over d^2}\right)^{1\over 2}\cos\Psi -{r_*\over d}\sin\Psi, &$\Psi_0>\arcsin{r_*\over d_0}$,}\nonumber}\\
\lefteqn{
\mu_2=\cases{
-\left(1-{r_*^2\over d^2}\right)^{1\over 2}\cos\Psi +{r_*\over d}\sin\Psi, &$\Psi_0<\upi-\arcsin{r_*\over d_0}$;\cr
1, &$\Psi_0\geq\upi-\arcsin{r_*\over d_0}$}\label{mu_s}
}
\end{eqnarray}
\citep{st08}. The condition for a \g-ray not to hit the star is $\Psi_0> \arcsin(r_*/d_0)$, but we can still use the presented formalism to calculate ${\rm d}\tau_{\gamma\gamma}/{\rm d}l$, needed for calculations of the distribution of the produced pairs. In the present coordinate system, $\mu$ is independent of $\phi$ (unlike the case in \citealt{dubus06}). For a given $\mu$, we can thus linearly integrate over $\phi$. The range of integration is $-\Phi_{\rm s}\leq\phi\leq\Phi_{\rm s}$, where $\Phi_{\rm s}$ equals \citep{bednarek97,st08}
\begin{equation}
\Phi_{\rm s}=\arccos\max\left[{(1-r_*^2/d^2)^{1/2}+\mu \cos\Psi\over (1-\mu^2)^{1/2}\sin\Psi},-1\right].
\label{phi_s}
\end{equation}
Thus, the integral over $\phi$ is replaced by $2\Phi_{\rm s}$. 

Given that formulae (\ref{mu_s}--\ref{phi_s}) depend on $\Psi$, it is convenient to integrate over it instead of $l$,
\begin{equation}
{\rm d}l={d_0\sin\Psi_0\over \sin^2\Psi}{\rm d}\Psi,
\label{dpsi}
\end{equation}
with which we write the final result as,
\begin{eqnarray}
\lefteqn{
\tau_{\gamma\gamma}(\epsilon, d_0,\Psi_0)=12 \sigma_{\rm T}d_0\left(m_{\rm e}c\over h \epsilon\right)^3\sin\Psi_0 \int_{\Psi_0}^\upi{{\rm d}\Psi\over \sin^2\Psi} \times
\nonumber}\\
\lefteqn{\quad
\int_{\mu_1(\Psi)}^{\mu_2(\Psi)}\!\!{\rm d}\mu{\Phi_{\rm s}(\Psi,\mu)\over (1-\mu)^2}\int_0^1\!\!{\rm d}\beta\,f_{\gamma\gamma}(\beta,\epsilon,\mu),\label{final}}\\
\lefteqn{
f_{\gamma\gamma}\equiv \beta {(3-\beta^4)\ln{1+\beta\over 1-\beta}-2\beta(2-\beta^2)\over (1-\beta^2)^3\left[\exp {2\over \epsilon\Theta_* (1-\mu)(1-\beta^2)}-1\right]},\label{fgg}}
\end{eqnarray}
except in the case of $\Psi_0=\upi$, where we keep integration over the length,
\begin{eqnarray}
\lefteqn{
\tau_{\gamma\gamma}(\epsilon, d_0,\upi)=
\label{taupi}}\\
\lefteqn{\quad
12\upi \sigma_{\rm T} r_* \left(m_{\rm e}c\over h \epsilon\right)^3 \int_{d_0/r_*}^\infty\!\!\!\!{\rm d}\delta \int_{(1-\delta^{-2})^{1/2}}^1\!\!{{\rm d}\mu\over (1-\mu)^2}
\int_0^1\!\!{\rm d}\beta\,f_{\gamma\gamma}(\beta,\epsilon,\mu).\nonumber
}
\end{eqnarray}
If we want to calculate ${\rm d}\tau_{\gamma\gamma}/{\rm d}l$ (needed for the distribution of the produced pairs), we obviously do not use the variable change (\ref{dpsi}).

When $d\gg r_*$, the blackbody emission can be approximated as coming from a point source. In that case, $\mu=-\cos\Psi$. Then, we need to integrate only over the \g-ray path and blackbody energies, see e.g., equation (A.9) of \citet{dubus06}. The optical depth is given by,
\begin{eqnarray}
\lefteqn{
\tau_{\gamma\gamma}(\epsilon, d_0,\Psi_0)=
\nonumber}\\
\lefteqn{\quad
{6\upi \sigma_{\rm T}r_*^2\over d_0 \sin\Psi_0} \left(m_{\rm e}c\over h \epsilon\right)^3 \int_{\Psi_0}^\upi{{\rm d}\Psi\over (1+\cos\Psi)^2}
\int_0^1\!\!{\rm d}\beta\,f_{\gamma\gamma}(\beta,\epsilon,\mu).
\label{point}}
\end{eqnarray}

In our case of emission from a jet perpendicular to the orbital plane and circular orbit, $d_0$ and $\Psi_0$ are related to the height along the jet, $z$, and the orbital phase, $\phi_{\rm b}$, by
\begin{equation}
d_0^2=z^2+a^2,\quad \cos\Psi_0= (a \sin i\cos\phi_{\rm b}-z \cos i)/d_0.
\label{d0Psi0}
\end{equation}

\label{lastpage}

\end{document}